\documentclass[a4paper]{jpconf}
\usepackage{graphicx}
\usepackage{cancel}
\usepackage{amsmath}
\usepackage{pgfplots}
\usepackage{overpic}
\usepackage{xcolor}
\usepackage{array}
\usepackage{caption}
\usepackage[backend=biber,style=numeric,sorting=none]{biblatex}
\usetikzlibrary{calc}
\usepgfplotslibrary{colormaps}
\usepackage{tikz}
\usetikzlibrary{arrows.meta}
\usepackage{cleveref}
\usepackage{subcaption}
\usepackage{todonotes} 

\usepackage[normalem]{ulem}

\usepackage{graphicx} 
\usepackage{xcolor}

\begin{document}
\title{Programming tension in 3D printed networks inspired by spiderwebs} 
%
\author{T.C.P. Masmeijer$^1$, C.C. Swain$^2$, J.R. Hill$^2$ and E. Habtour$^1$}
\address{$^1$ Aeronautics \& Astronautics, The University of Washington, Seattle, WA 98195, USA}
\address{$^2$ Department of Mechanical Engineering,
Brigham Young University, Provo, UT 84604, USA}
\ead{habtour@uw.edu}
\date{February 2025}
\section*{Abstract}

Each element in tensioned structural networks---such as tensegrity, architectural fabrics, or medical braces/meshes---requires a specific tension level to achieve and maintain the desired shape, stability, and compliance. These structures are challenging to manufacture, 3D print, or assemble because flattening the network during fabrication introduces multiplicative inaccuracies in the network's final tension gradients. This study overcomes this challenge by offering a fabrication algorithm for direct 3D printing of such networks with programmed tension gradients, an approach analogous to the spinning of spiderwebs. 
The algorithm: (i) defines the desired network and prescribes its tension gradients using the force density method; (ii) converts the network into an unstretched counterpart by numerically optimizing vertex locations toward target element lengths and converting straight elements into arcs to resolve any remaining error; and (iii) decomposes the network into printable toolpaths; Optional additional steps are: (iv) flattening curved 2D networks or 3D networks to ensure 3D printing compatibility; and (v) automatically resolving any unwanted crossings introduced by the flattening process.
The proposed method is experimentally validated using 2D unit cells of viscoelastic filaments, where accurate tension gradients are achieved with an average element strain error of less than 1.0\%. The method remains effective for networks with element minimum length and maximum stress of 5.8 mm and 7.3 MPa, respectively. The method is used to demonstrate the fabrication of three complex cases: a flat spiderweb, a curved mesh, and a tensegrity system. The programmable tension gradient algorithm can be utilized to produce compact, integrated cable networks, enabling novel applications such as moment-exerting structures in medical braces and splints.

\textbf{Keywords:} Tensioned structures, Fabrics, Tensegrity, Programmed tension, Medical braces, Bioinspired 

\begin{figure}[htbp]
    \centering
    \includegraphics[width=.9\textwidth]{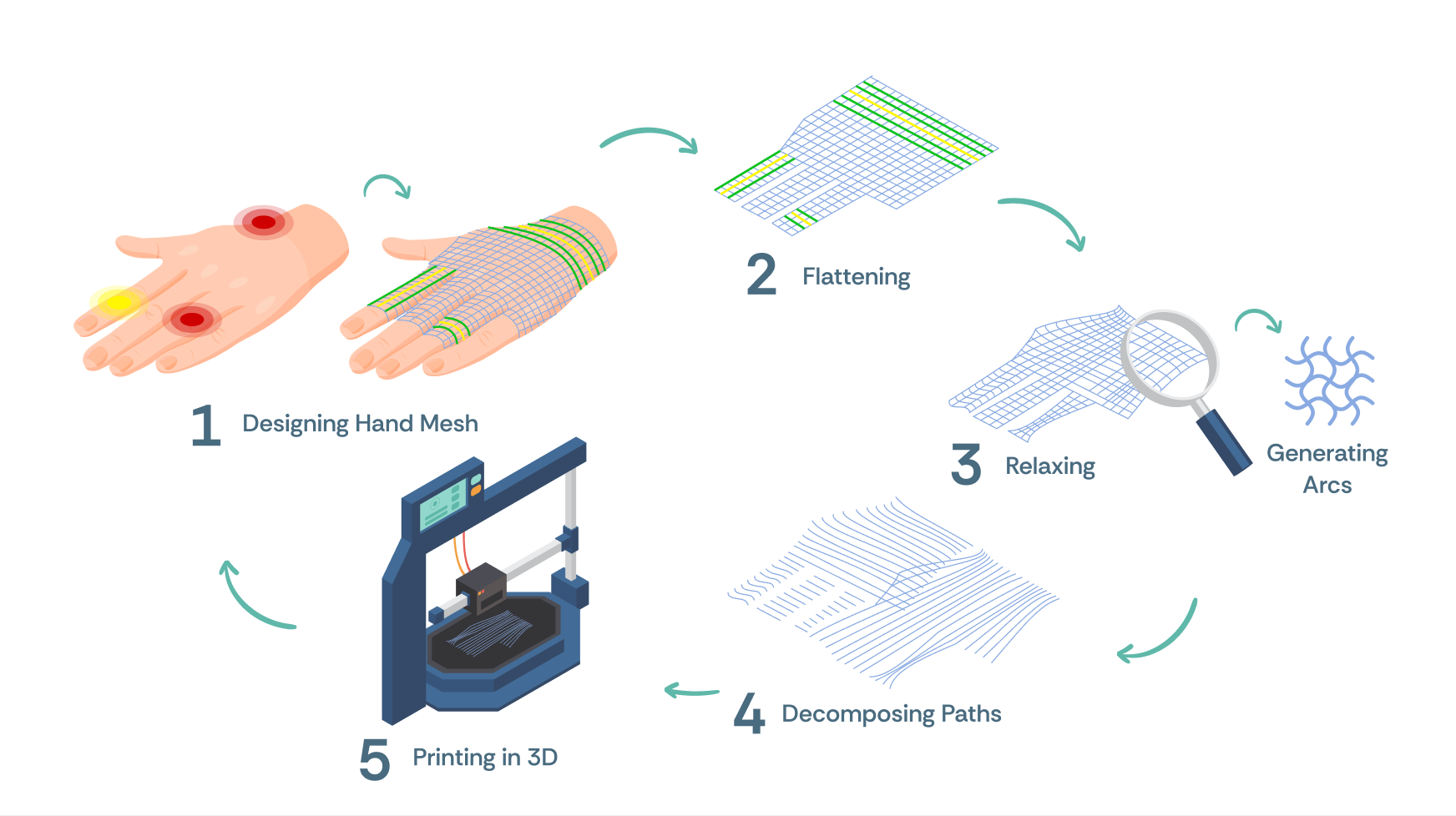}
    \caption{\textbf{Graphical abstract}: An example application of a 3D printed cable network with programmable tension gradients is a custom, patient-specific, compression cast. The steps to automatically manufacture them consist of
1) Designing a cable network with a tension gradient;
2) Flattening it;
3) Finding the networks unstreteched counterpart that would result in the target design and tension gradient when the patient wears the network;
4) Decomposing the network into continuous paths;
And 5) 3D printing the structure.
%
 }
    \label{fig:Graphical-abstract}
\end{figure}
\section{Introduction}
In any tensioned membrane structural system, precise tension levels between individual elements are critical to achieving and maintaining the desired stability, compliance, and geometry. While historically applied in civil engineering contexts such as lightweight catenary systems and tensioned fabric roofs \cite{pollini_gradient-based_2021, klosowska_self-stress_2018, gilewski_influence_2017}, these systems now demonstrate significant potential across emerging domains. For example, recent advances include assistive wearables for medical rehabilitation \cite{lessard_crux_2017}, enhanced actuation in soft robots \cite{chen_towards_2024, renda_dynamic_2014}, tensegrity network for lightweight  robots \cite{layer_control_2024,zappetti_variable-stiffness_2020, brown_development_2024, shekastehband_effects_2017}, and adaptive systems for space debris removal \cite{Furuya1992, Shinde2021, Shang2025}. In all applications, cable tension distribution is fundamental, not only for maintaining structural integrity but also for defining dynamic behaviors, such as natural frequencies and deflection modes \cite{pajunen_design_2019, zhang_initial_2017}. Nevertheless, achieving a precise tension distribution in such structures remains a persistent challenge due to two main engineering limitations; (1) tension often only emerges after the application of external loads, rather than being inherent in the initial fabrication, and (2) conventional fabrication methods are inefficient and may fail to produce consistent or optimal prestress states \cite{chen_initial_2012, kim_estimation_2007}. These limitations are especially pronounced in miniaturized or thin systems, where small deviations in prestress can cause unintended substantial variations in mechanical performance. Inspired by the tensioning process in spiderwebs, the authors developed and experimentally validated an algorithm for engineering self-tensioning elements within 3D-printed membrane/mesh networks that overcomes these limitations. 

\subsection{Challenges and recent advancements}
Despite advances in optimization techniques for defining the tension of structural elements, realizing the intended tension state during manufacturing remains technically difficult \cite{denning_tuning_2024}. This is because construction tends to require jigs, multiple people, or time-consuming individual component assembly. Assembly solutions can limit passive cable control or require active cable control \cite{Rhodes2019}. Additive manufacturing has offered a promising route to fabricating small, prestressed structures and has been used to produce tensegrity lattices with tunable band gaps and dynamic properties \cite{Pajunen2021, lee_3d-printed_2020}. Parallel developments in 3D knitting further enabled localized tailoring of mechanical responses by varying material composition and patterning \cite{Luo2022, Sanchez2023}. While notable progress has been made in 3D printing and 3D knitting, both approaches exhibited inherent limitations. In current fully automated 3D printed tensegrities, designed tension states only emerged after an external load was applied \cite{lee_3d-printed_2020, Pajunen2021, Sabouni2024}. These automated methods aimed to overcome the labor-intensive nature of piecewise fabrication, where cables and bars were assembled separately. However, the as-printed structures were typically unstressed. In contrast, 3D knitting allowed fully automated fabrication of complex, deformable structures, but required equipment that was more costly than widely available 3D printers. Several efforts have explored 3D printing prestressed structures using Fused Filament Fabrication (FFF), including strategies that involve post-processing to remove sacrificial molds \cite{pajunen_design_2019, Lee2020, Sabouni2024}, but a scalable method to directly manufacture tension-programmed networks remains a persistent challenge.

\subsection{Why spiderwebs-inspiration?}
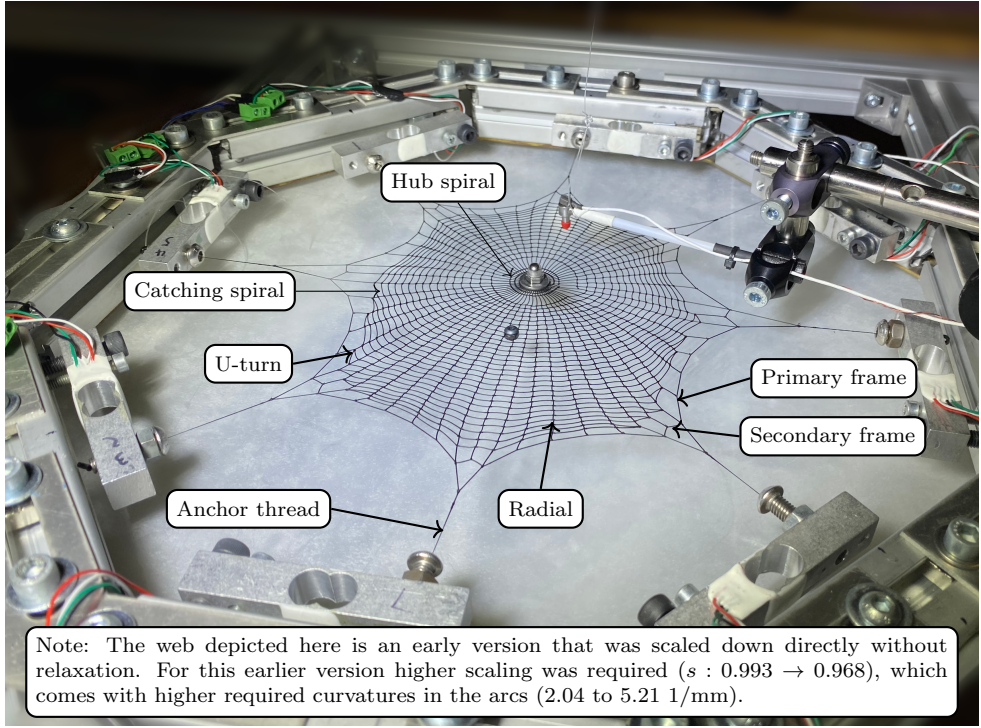
\begin{figure}
    \centering
    \begin{tikzpicture}
        \node[anchor=south west, inner sep=0pt] (image) at (0,0)
            {\includegraphics[width=.8\textwidth]{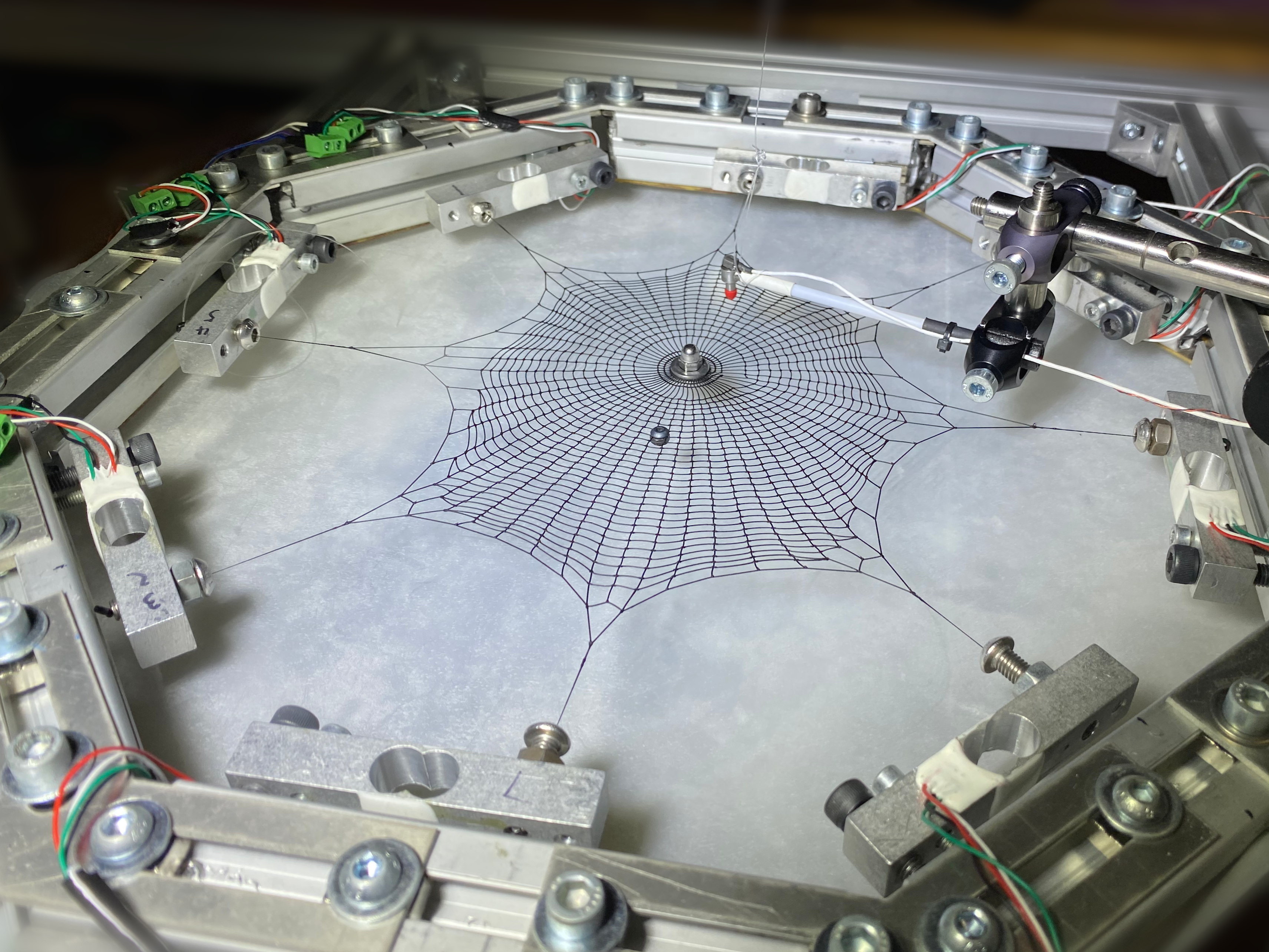}};
        
        \begin{scope}[x={(image.south east)}, y={(image.north west)}]
            
            \node[draw, fill=white, text=black, thick, font=\scriptsize, rounded corners, minimum size=0.5cm] (hub_s) at (0.45,0.75) {Hub spiral};
            \draw[->, thin] (hub_s) -- (0.52,0.62);
            
            \node[draw, fill=white, text=black, thick, font=\scriptsize, rounded corners, minimum size=0.5cm] (spiral) at (0.21,0.6) {Catching spiral};
            \draw[->, thin] (spiral.east) -- (0.385,0.6);
            
            \node[draw, fill=white, text=black, thick, font=\scriptsize, rounded corners, minimum size=0.5cm] (radial) at (0.55,0.3) {Radial};
            \draw[->, thick] (radial.north) -- (0.562,0.42);
            
            \node[draw, fill=white, text=black, thick, font=\scriptsize, rounded corners, minimum size=0.5cm] (Uturn) at (0.25,0.5) {U-turn};
            \draw[->, thick] (Uturn.east) -- (0.357,0.515);
            
            \node[draw, fill=white, text=black, thick, font=\scriptsize, rounded corners, minimum size=0.5cm] (frame1) at (0.851,0.48) {Primary frame};
            \draw[->, thick] (frame1.west) -- (0.69,0.45);
            
            \node[draw, fill=white, text=black, thick, font=\scriptsize, rounded corners, minimum 
            size=0.5cm] (frame2) at (0.85,0.4) {Secondary frame};
            \draw[->, thick] (frame2.west) -- (0.685,0.411);

            \node[draw, fill=white, text=black, thick, font=\scriptsize, rounded corners, minimum 
            size=0.5cm] (anchor) at (0.25,0.3) {Anchor thread};
            \draw[->, thick] (anchor.east) -- (0.45,0.27);
            \node[draw, fill=white, text=black, thick, font=\scriptsize, rounded corners,
              text width=0.75\textwidth, align=justify] 
              at (0.5,0.075) {Note: The web depicted here is an early version that was scaled down directly without relaxation. For this earlier version higher scaling was required ($s:0.993 \rightarrow 0.968$), which comes with higher required curvatures in the arcs (2.04 to 5.21 1/mm).};
        \end{scope}
    \end{tikzpicture}
    \caption{A 3D-printed spiderweb-like structure suspended in a frame. The web incorporates design features commonly observed in spider orb webs. These features are annotated in the figure according to the nomenclature reported by Zschokke et al. \cite{Zschokke1999}.}
    \label{fig:spiderweb_photo}
\end{figure}

Spiderwebs are natural networks designed for tunable tension gradients. For instance,  the typical tension ratio in an orb web between its anchor threads, frame threads, and radii is approximately 10:7:1 \cite{Wirth1992},  see \cref{fig:spiderweb_photo}. Spiders build these structures in a matter of hours by sequentially laying down continuous elements such as bridging threads, primary/secondary frames, radii, sticky spirals, and a central hub \cite{Zschokke1996, Ramousse1976}. This remarkable fabrication approach of a tensioned structure is created seamlessly, with the spider constantly maintaining the desired tensions as it builds. While spiderwebs are built strand by strand under tension, the networks we propose are fabricated in an unstressed state and only take on their designed tensioned shape upon application. Still, they are similarly constructed as a continuous path--- making them well-suited for extrusion-based 3D printing.

\subsection{Approach}
In this study, we presented a method for manufacturing network structures with programmable tension gradients using accessible 3D printing, thereby overcoming the limitations mentioned above. The approach is also explained graphically in \cref{fig:Graphical-abstract}. Desired tension networks, for example, were designed using the Force Density Method (FDM). Although 3D printing a network in its equilibrium shape produced the correct geometry, it would not achieve the intended tension distribution. Therefore, the network was converted into an unstressed counterpart that attained the desired shape and tension after assembly. In order to manufacture systems similar to orb spider-webs, the network was decomposed into continuous printable paths. The approach was not limited to planner networks such as spider webs; procedures were also provided to flatten curved 2D (2.5D) or fully 3D structures.
The novelty of our method is reported in this study, which is fundamentally different from 3D knitting, as programmable deformation was achieved by tuning the tension in individual edges rather than by adjusting local material properties.

\subsection{Impact}
Our tension programmable method for structural networks opens the door to impactful, real-world applications. One of the clearest opportunities lies in the design of compression casts or splints. Traditional materials like plaster and fiberglass are rigid and nonadjustable, often resulting in excessive pressure that requires valving \cite{Crickard2011} or complete recasting \cite{Halanski2008}. In contrast, our approach enables compression to be actively and locally tuned by embedding programmable tension gradients during the fabrication process. When combined with a ratcheting system or control cable, it becomes possible to design splints that adapt to the patient over time--- improving comfort and reducing the need for medical intervention. Because our method is compatible with low-cost 3D printers, these devices can be manufactured locally, in homes or community spaces, enabling affordable and patient-specific care.

A second application area is wearable robotics. Devices like cable-driven exosuits (e.g., CAREX \cite{mao_design_2012} and CRUX \cite{lessard_crux_2017}) depend on tensioned elements to transmit forces and assist movement. Our method enables these elements to be directly encoded with programmable stiffness and directional force transfer, thereby reducing complexity and allowing for more compact and lightweight designs. The ability to print such structures without bulky knitting machines further expands access to research, prototyping, and personalized solutions in rehabilitation and assistive technologies.

Compared to fully automated methods, our approach involves more manual effort, as bars must be printed separately from cables and combined afterwards. However, it significantly reduces overall assembly time compared to traditional piecewise fabrication as the cables can be printed as one piece, and, crucially, it embeds the intended tension state directly into the structure, eliminating the need for external loading and final assembly adjustments. This makes our method uniquely suited for rapid and accurate prototyping of functional tensegrity systems.

This article is structured as follows. In \cref{sec:Results}, the applicability of the method to three cases is detailed; namely,  a 2D spiderweb-inspired network (\cref{sec:spider web}),  a 2.5D moment-exerting mesh (\cref{sec:moment exerting mesh}), and a 3D tensegrity system (\cref{sec:tensegrity structure}). Section~\ref{sec:Methods} . The method's steps for designing, processing, and manufacturing network structures are described. A Python code accompanying this paper is publicly available to reproduce the results and test its applicability to other networks \cite{masmeijer2025_PNS}.
The method's experimental validation and limitations are described in \cref{sec:validation}. The discussion of results and conclusions are provided in\cref{sec:conclusions}. 
%
%
\section{Results}\label{sec:Results}
In this section, we demonstrate three cases of structural networks manufactured using the proposed method. Case-1 was an orb spiderweb to illustrate nature-tensioning in a 2D network. Case-2 was a moment-exerting mesh for a medical arm compression, which was a 2.5 network. Finally, Case-3 was a 3D tensegrity network to demonstrate the applicability of our method to a complex structural system.

\subsection{Case-1: Spider web} \label{sec:spider web}
\begin{figure}[h]
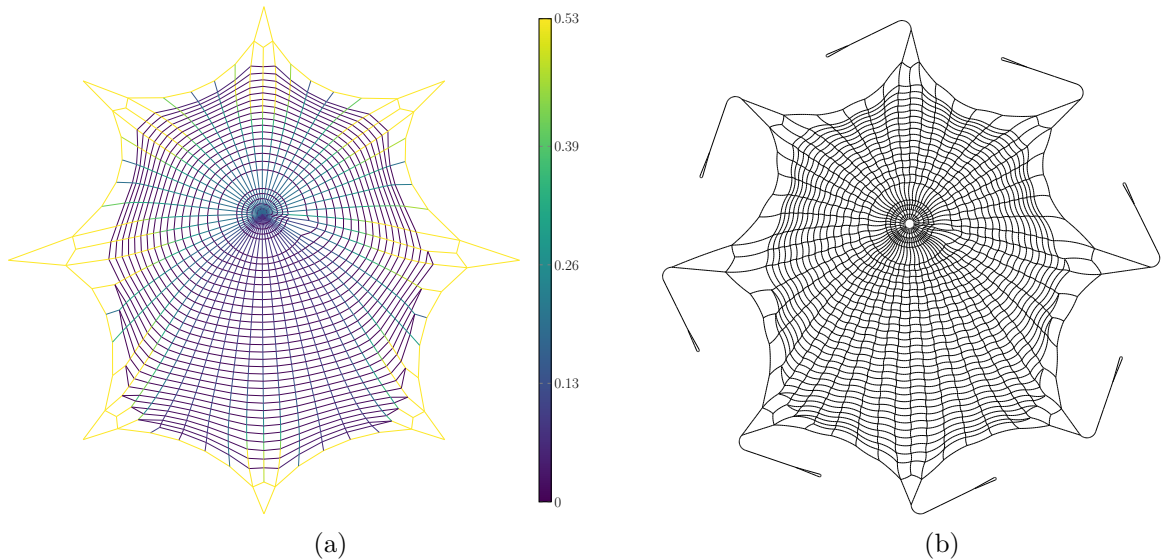

    \centering
    \begin{subfigure}{0.53\textwidth}
    \resizebox{0.9\textwidth}{!}{%
        \input{tikzfigures/tikz_image_web}
    }
    \caption{}
    \label{fig:spider-forces}
    \end{subfigure}
    \begin{subfigure}{0.46\textwidth}
    \resizebox{0.9\textwidth}{!}{%
        \input{tikzfigures/tikz_image_web_arc2}
    }
    \caption{}
    \label{fig:spider-arcs}
    \end{subfigure}
    \caption{a) A spiderweb-like network is designed using the Force Density Method. The color of edges indicates the tension in Newton. b) An unstretched counterpart of the web with arcs before 3D printing. Post-processing involves adding eight connection loops and manually modifying a hole at the center to prevent material aggregation.}
    \label{fig:spider-web-both}
\end{figure}

A spider web was the source of our inspiration due to its ingenious tension gradient approach for realizing its remarkable and highly specialized structural network. Using the proposed method, spiderweb-like structures were fabricated with controlled tension distributions that closely mimic those of real webs. The structure presented here follows average design variability as measured by Vollrath et al.\ and Rhisiart et al.\ \cite{Rhisiart1994, Vollrath1997}, with modeled tension gradients based on measurements by Wirth et al.\ \cite{Wirth1992}. The designed orb web with tension gradient and its unstretched printable counterpart are depicted in \cref{fig:spider-web-both}.

The target geometry was initially generated using the force density method (see \cref{sec:FDM}). Converting the target geometry into an unstretched counterpart took three steps, which are detailed in \cref{sec:net_opt}: Relaxing the vertexes, scaling down the web and converting edges into arcs with target arc lengths. The final network was decomposed into continuous paths that mimic the building steps of real spider webs: a central hub spiral, a catching spiral with U-turn, a primary frame, eight secondary frame sections, and eight anchor threads, please refer to \cref{fig:spiderweb_photo} for descriptions of web sections. Methods for decomposing networks into continuous paths are detailed in \cref{sec:printable paths}. The 3D printed result, suspended in a customized frame, is shown in \cref{fig:spiderweb_photo}. Optimization convergence was achieved after 1852 iterations with a final error, $\epsilon_{1852}$, of approximately 0.153 (using a damping parameter $\beta = 0.1$ and an optimization convergence tolerance $\tau = 10^{-6}$). The web was scaled by a factor $s = 0.993$.

The successful printing of this web-like structure demonstrated that complex, highly connected networks could be constructed directly and automatically using our method. The capability of integrating tension gradients into spider web-like structures distinguished our approach from experimental studies on artificial spider webs \cite{Su2021, Buehler2021}, which focused on geometry and material properties but did not include appropriate tensions. Quantitative validation was performed using simple unit cells (\cref{sec:validation}).

\subsection{Case-2: Moment-exerting mesh} \label{sec:moment exerting mesh}
Tension gradients in a network can be designed to deform and exert loads in desirable ways. To demonstrate this, a moment-exerting mesh of a thin-walled cylinder was presented. The surface of the thin-walled cylinder under bending experiences stress can be described by
\begin{equation}
    \sigma_{\text{bending}} = \frac{M R \cos\theta}{I},
    \label{eq:bending_stress}
\end{equation}
where $M$ is the bending moment, $R$ the radius, $\theta$ the circumferential angle, and $I$ is the second moment of area.

\begin{figure}
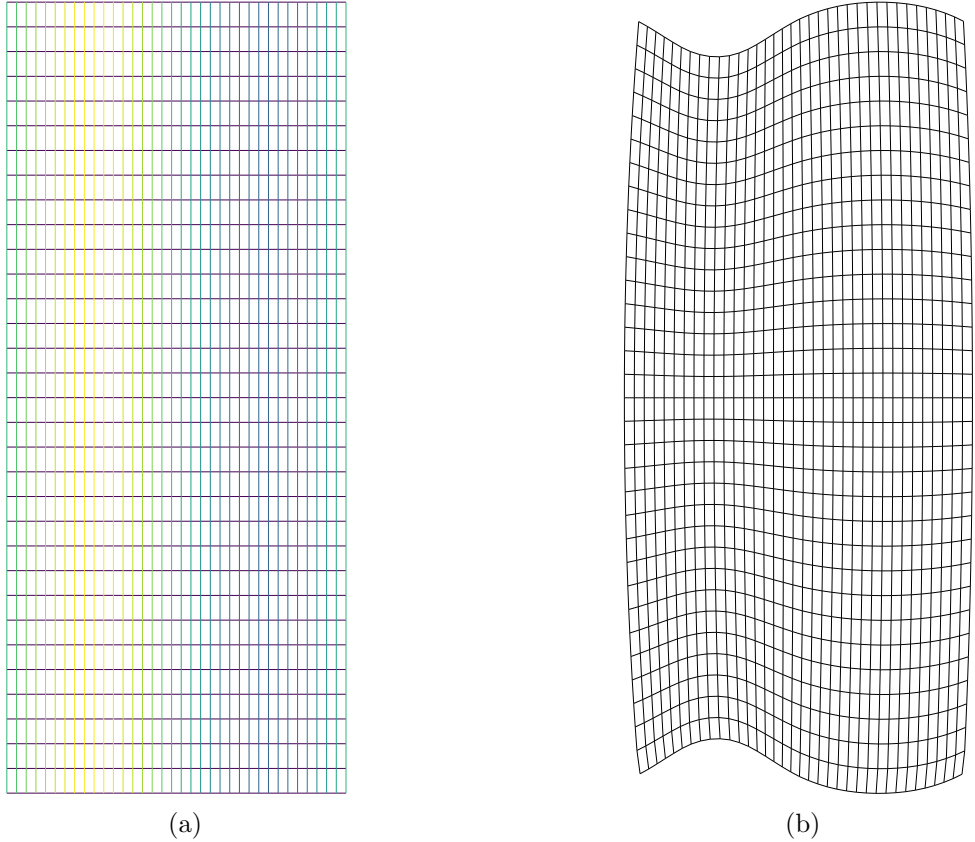

    \centering
    \begin{subfigure}{0.49\textwidth}
        \centering
        \rotatebox{0}{ 
            \resizebox{.6\textwidth}{!}{%
                \input{tikzfigures/tikz_balloonbender0}
            }
        }
        \caption{}
        \label{fig:balloon-bender-0}
    \end{subfigure}
    \hspace{0.002\textwidth}
    \begin{subfigure}{0.49\textwidth}
        \centering
        \rotatebox{0}{ 
            \resizebox{.6\textwidth}{!}{%
                \input{tikzfigures/tikz_balloonbender1}
            }
        }
    \caption{}
        \label{fig:balloon-bender-1}
    \end{subfigure}
    \caption{Moment exerting wrap. a) The tension gradient of a cylinder unwrapped over the network, and b) the network after optimizing the edge lengths.}
    \label{fig:balloon-benders}
\end{figure}

To replicate this stress distribution, the cylinder was unwrapped into a flat sheet and discretized using $n_x = 35$ vertical and $n_y = 32$ horizontal members, as shown in \cref{fig:balloon-benders}. The force density of the vertical edges was varied as a function of $\theta$ using
\begin{equation}
    q(\theta) = q^0 + \Delta q \sin(\theta), \qquad 
    q_i = q(\theta_i), \qquad 
    \theta_i = \frac{2\pi i}{n_x}, \quad i=0,\dots,n_x-1.
\end{equation}
with $q^0 = 0.16$ N/mm, $\Delta q = 0.076$ N/mm. Horizontal elements were assigned a constant force density of 0.035 N/mm. Boundary vertices were fixed. The resulting network is shown in \cref{fig:balloon-bender-0}. To ensure sufficient stiffness, vertical members were printed with three layers of filament. The designed mesh was relaxed, scaled down and edges were turned into arcs, as detailed in \cref{sec:net_opt}. The optimized shape is shown in \cref{fig:balloon-bender-1}. The mesh was relaxed using a damping $\beta = 0.7$ and convergence tolerance $\tau = 10^{-6}$. After 1,703 iterations, the shape converged with an error of $\epsilon_{1703} = 0.017$ and the mesh was scaled down with a scaling factor $s = 0.996$.

\begin{figure}[h]
    \centering
    \begin{minipage}{0.05\textwidth}
        \subcaption{}\label{fig:wrap-a}
    \end{minipage}%
    \begin{minipage}{0.9\textwidth}
        \includegraphics[width=\linewidth]{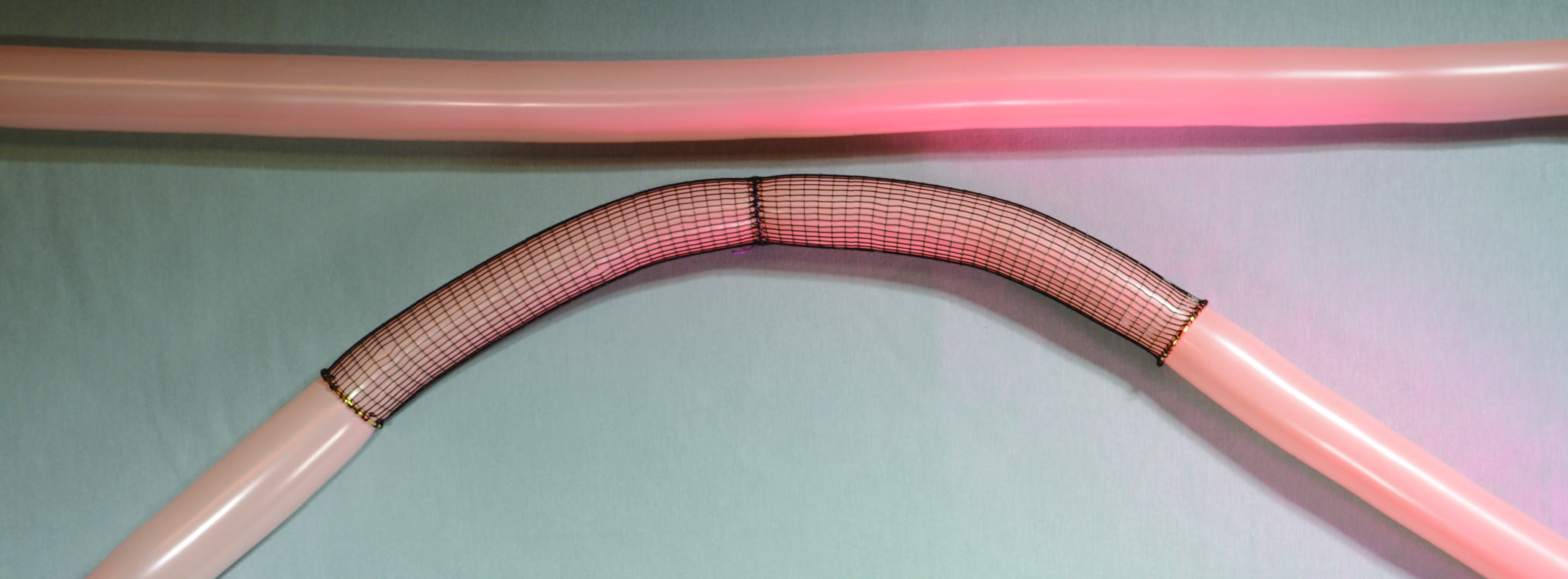}
    \end{minipage}

    \vspace{0.5em}

    \begin{minipage}{0.05\textwidth}
        \subcaption{}\label{fig:wrap-b}
    \end{minipage}%
    \begin{minipage}{0.9\textwidth}
        \includegraphics[width=\linewidth]{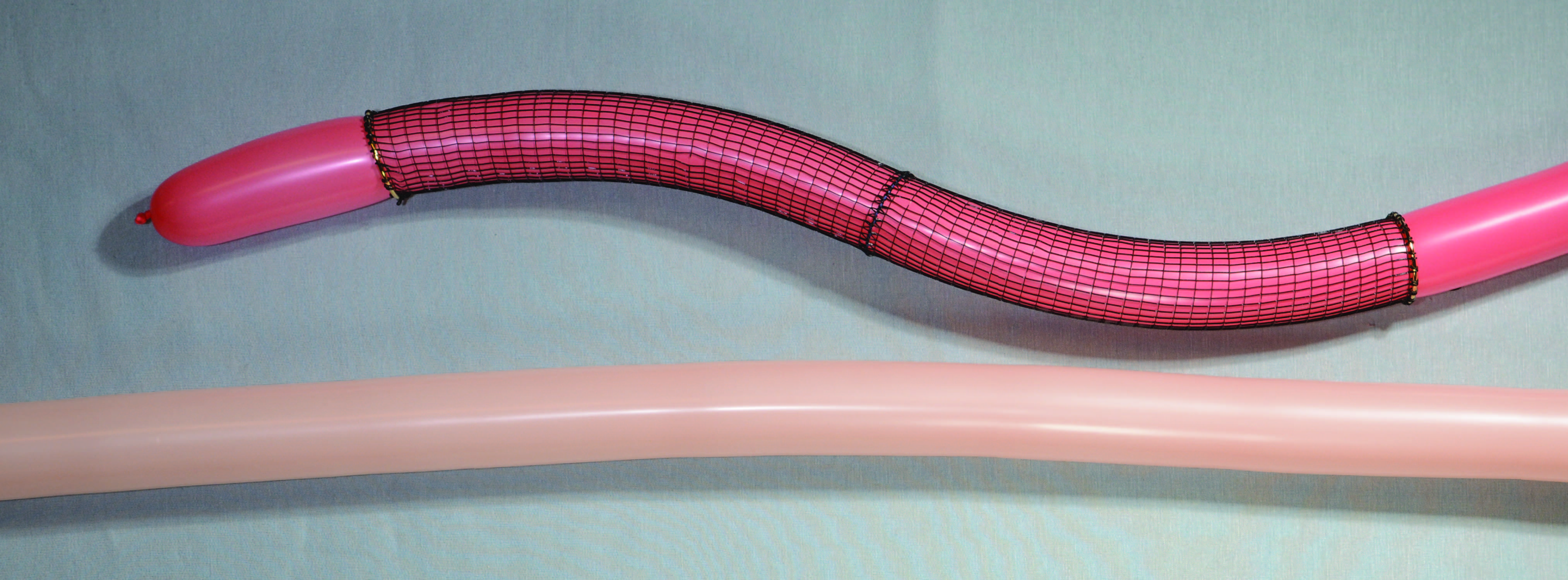}
    \end{minipage}

    \caption{Moment exerting wraps on balloons next to unloaded balloons. 
    (a) Two wraps in the same orientation, resulting in a U-shape. 
    (b) Two wraps in opposing directions, resulting in an S-shape.}
    \label{fig:moment-wrap-image}
\end{figure}

To enable physical testing, loops were added at the top and bottom of each vertical strand, allowing a metal ring to pass through. A wire was used to connect the left and right vertical boundaries. When applied to a balloon, the wrap produced a directional moment: two wraps in the same direction generated a U-shape, while opposing directions yield an S-shape (see \cref{fig:moment-wrap-image}).

This approach demonstrated how stress gradients can be embedded into a 3D printed network to control out-of-plane deformation. Parallel developments in 3D knitting have shown that local patterning can tune the mechanical response in soft robots and assistive gloves \cite{Luo2022, Sanchez2023}. Our method achieved a similar outcome, but through force-based optimization and filament deposition, offering a distinct working mechanism. Potential applications include customizable compression casts or splints with locally adjustable tension \cite{Crickard2011, Halanski2008}.

\subsection{Case-3: Tensegrity system} \label{sec:tensegrity structure}
Printing 3D cable networks suspended in the air is not feasible using planar 3D printing techniques, such as FFF. Instead, the networks must first be flattened while preserving their topology. This is particularly challenging when the network is spatially complex and densely connected, as is the case in tensegrity structures. Flattening can introduce internal crossings between edges that were not originally intersecting in 3D space. These crossings must be identified and resolved to preserve the intended topology (see \cref{sec:flattening} for details). Note that the flattening and crossings resolution methods are automated to ensure the methods scale to more complicated networks.

To showcase this challenge, a classic tensegrity structure is manufactured: the expandable octahedron \cite{Pugh1976-cw}. This structure includes an intricate 3D cable network, making it a good candidate for validating the flattening and intersection-resolution methods. It consists of three parallel pairs of struts connected by 24 cables, forming a symmetric, force-balanced configuration.

\begin{figure}[h!]
    \centering
    \begin{minipage}{0.45\linewidth}
        \centering
        \includegraphics[width=\linewidth]{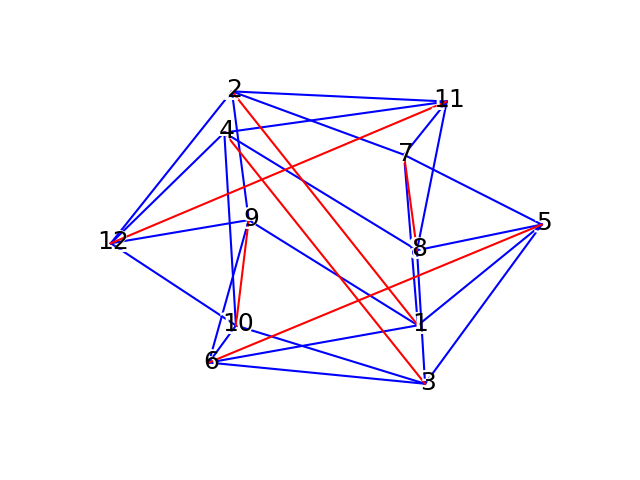}
        \subcaption{}
        \label{fig:tensegrity_design}
    \end{minipage} \hfill
    \begin{minipage}{0.45\linewidth}
        \centering
        \includegraphics[width=\linewidth]{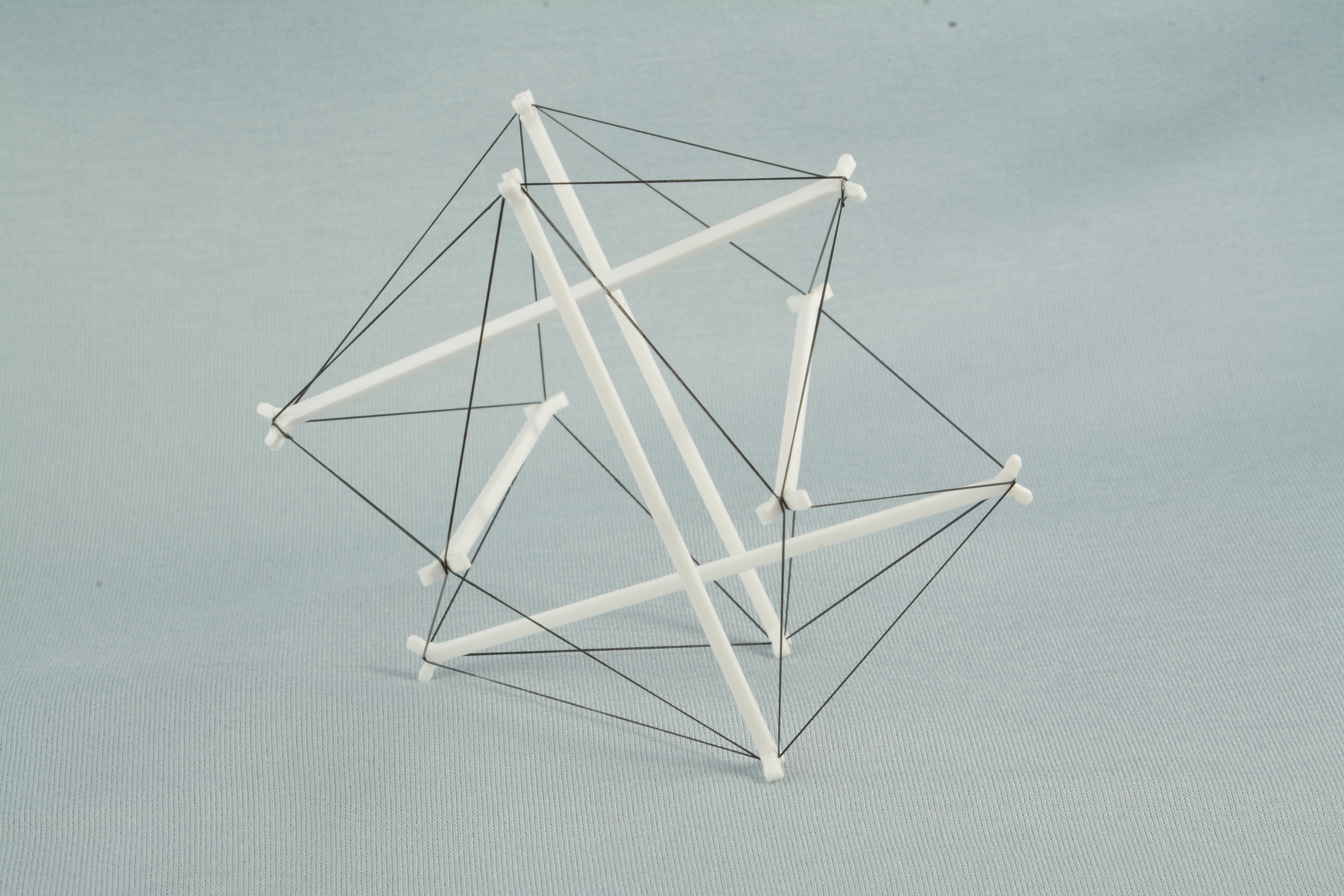}
        \subcaption{}
        \label{fig:tensegrity_printed}
    \end{minipage}
    \caption{a) Graphical image of the expandable octahedron (red: struts, blue: cables) and b) the assembled tensegrity.}
    \label{fig:tensegrity_comparison}
\end{figure}

The force density method is a useful tool for designing tensegrity structures, but not every set of force densities results in a stable configuration. Previous work has shown that the force density ratio $-q_{\text{strut}} = 1.5 q_{\text{cable}}$ did not result in a stable system where each cable and each strut has equal lengths, and the length ratio between struts and cables was 1.632 \cite{Tibert2003, Koohestani2012, Tran2010}. The designed structure is depicted in \cref{fig:tensegrity_design} (an animation is available in electronic viewings). For a strut length of 80 mm and a cable stress of 7.5 MPa ($\approx20\%$ strain), the resulting force densities were -0.018 for struts and 0.012 for cables.

The tensegrity was fabricated in two steps: struts were printed using PETG and the cable network from TPU. Due to the significantly higher stiffness of PETG and the higher cross-sectional area of the struts with respect to the cables, the strut deformation was neglected: $l^0 = l^1$.

The cable network was flattened using a polar coordinate transformation, and intersecting edges were resolved before optimization. The designed and fabricated tensegrity networks are shown in \cref{fig:tensegrity_comparison}. Flattening the cable network had the side effect of introducing additional crossings. These additional crossings were automatically resolved using the methods described in \cref{sec:flattening}.

Optimization was performed with parameters $\beta = 0.1$, $\tau = 10^{-6}$, yielding a final error $\epsilon_{1907} = 1.7 \times 10^{-4}$. The network was scaled down with a ratio $s = 0.9999$.

Programmable tensegrity structures such as this could support applications in cable-driven exosuits, such as CAREX \cite{mao_design_2012} or CRUX \cite{lessard_crux_2017}, where tunable stiffness and force transmission are critical.

\section{Methods} \label{sec:Methods}
This section discusses methods for manufacturing 3D printed networks with programmable tensions. We detail the methods for  (i) defining the goal network using the force density method (ii) converting this goal network into an unstretched counterpart; (iii) transforming the network into a 3D printing compatible structure; (iv) flattening 2.5D/3D networks; and (v) accounting for unwanted crossings. Steps (i) and (ii) are also described graphically in \cref{fig:design-validation-method}, step (iii) in \cref{fig: unit-cell}, and steps (iv) and (v) in \cref{fig:tensegrity_networks}.
\subsection{Form-finding}\label{sec:FDM}
Tensioned structural networks are typically defined using a form-finding technique: the force density method \cite{Schek1974}. This technique was developed to design tensioned networks with equilibrated structures for given boundaries and topologies \cite{pauletti2020outline}, such as tensioned roofs (e.g. the Munich Olympic Stadium roof (1972) \cite{Schek1974} and fabric formwork systems for the construction of curved concrete shells \cite{Echenagucia2019}). The force density method linearizes the static force equilibrium equations, enabling the fast form-finding of networks with desired shapes and tension gradients. First, the topology of a network containing $M$ edges and $N$ vertices are described by an $M\times N$ connectivity matrix $C_s$:
\begin{equation}
C_s(i, j)=\left\{\begin{array}{cl}
+1 & \text { if edge i starts at vertex j } \\
-1 & \text { if edge i ends at vertex j } \\
0 & \text { in the other cases}.
\end{array}\right.
\end{equation}
Next, a force density $q_i$ is assigned to each $i$-th edge in the network, where $q_i$ represents the ratio of tension $F_i$ to stretched length $l_i$ of the edge, as $q_i = F_i/l_i$. Vertices are divided into free $\textbf{x}$ and fixed $\textbf{x}_f$ vectors. Optionally, an external load vector $\textbf{p}$ can be applied at $\textbf{x}$. The free vertex coordinates $\textbf{x}$ can be found by solving
\begin{align}\label{eq:FD-method}
    \textbf{x} &= D^{-1}\left(\textbf{p} - D_f\textbf{x}_f \right) {\text{, with}} \\
    D &= C^TQC {\text{, and}} \\
    D_f &= C^TQC_f.
\end{align}
Where $Q$ represents the diagonal matrix of force densities $q_i$. The full connectivity matrix is divided into free and fixed columns according to $C_s = [C\quad C_f]$. The tension in each edge can be determined using $F_i=l_iq_i$. Optionally, a user can update the force densities in a network until the desired shape and tension gradient are achieved. If required, a nonlinear force density method can be used to incorporate constraints on the network's node locations, tensions, edge lengths, or any desired combination. A network that meets these constraints is found by employing numerical optimization \cite{Schek1974}. The force density method does not require a constitutive model for form-finding. The unstretched lengths of the edges, $l^0$, are determined only after form finding, based on the engineering strain relationship and the general stress definition. These quantities are computed independently for each edge:
\begin{align}
l_i^0 &= \frac{l_i}{1 + \epsilon(\sigma)}, \quad \text{where} \quad \sigma_i = \frac{F_i}{A_i}.
\end{align}
This calculation requires a constitutive model, $\epsilon(\sigma):=f(\sigma)$, and the cross-sectional areas $A$ of the edges. An Ogden hyperelastic material model is constructed from tensile tests, and the cross-sectional area is determined by printing a 36 m continuous fiber and measuring the amount of filament used (more info in \ref{AP:material model}). 


\subsection{Network optimization, scaling, and arc generation} \label{sec:net_opt}
Printing the equilibrated form directly would result in the correct geometry, but without tensions. To ensure the designed tension gradient, each edge in the network must be manufactured with the unstretched lengths \( l^0 \). The first step towards achieving this is by numerically optimizing the vertex locations. 

The optimization problem is highly coupled since adjusting the coordinates of a vertex to satisfy one length constraint inevitably affects the lengths of neighboring edges. This coupling makes the optimization procedure challenging, as a coordinate update in one iteration can propagate errors into subsequent ones, leading to oscillations or slow convergence. Common state-of-the-art optimization techniques, such as L-BFGS-B \cite{Zhu1997}, update all coordinates simultaneously based on global gradient and Hessian approximations, which can result in conflicting updates across coupled regions. These methods often produce exact but computationally intensive steps, making them relatively inefficient for problems with strong local dependencies. In contrast, a Gauss-Seidel optimization algorithm--- often referred to as a relaxation method ---updates a single vertex pair at a time and immediately incorporates each change into subsequent corrections, promoting more stable and efficient convergence. Moreover, because these updates are localized, the algorithm scales linearly with network size in sparse systems, making it well-suited for large-scale form-finding tasks \cite{Legarra2008}.

An overview of the Gauss-Seidel optimization algorithm is summarized as follows:
\begin{enumerate}
\item Use the form retrieved with the force density method as the initial guess \( \mathbf{x}_0 \). For curved or 3D networks, the initial guess first requires flattening and possibly intersection-resolution, as detailed in \cref{sec:flattening}.
\item Iterate over each vertex pair in the network and update their coordinates $\mathbf{x}^a$ and $\mathbf{x}^b$. The update procedure for a generic edge is
\begin{equation}\label{eq:Gauss-Seidel-update}
\begin{aligned}
    \mathbf{x}_{k+1}^a &= \mathbf{x}_k^a + \beta (l^0 - l_k^1) \frac{\mathbf{x}_k^a - \mathbf{x}_k^b}{2 l_k^1}, \\
    \mathbf{x}_{k+1}^b &= \mathbf{x}_k^b - \beta (l^0 - l_k^1) \frac{\mathbf{x}_k^a - \mathbf{x}_k^b}{2 l_k^1}.
\end{aligned}
\end{equation}
Here $l_k^1 = \|\mathbf{x}_k^a - \mathbf{x}_k^b\|$ is the current edge length in iteration $k$, and $\beta \in (0,1]$ is a numerical damping factor to improve convergence (e.g., $\beta = 0.1$).
\item Repeat step (ii) until convergence or until a maximum number of steps is reached. Convergence is reached when the difference between the current total error \( \epsilon_k \) and the previous step’s error \( \epsilon_{k-1} \) is smaller than a predefined tolerance \( \tau \), e.g., \( 10^{-6} \).
\end{enumerate}

In many optimization problems, care is taken to avoid convergence to local minima in favor of finding a global minimum. This is typically addressed by exploring a variety of initial conditions or applying regularization techniques. However, in our case, the nearest local minimum is preferred. While a lower-error configuration could, in theory, be found by allowing a vertex to move beyond its neighboring vertices, such a result would introduce edge crossings and distort the initial geometry. This would lead to forms that are difficult or impossible to manufacture. To preserve manufacturability, the optimization is constrained to maintain the original geometric layout. Therefore, a low damping factor $\beta$ is recommended to ensure that vertex updates are conservative and do not lead to large, destabilizing changes in geometry. The importance of the initial form $\textbf{x}^0$, which influences the local minimum reached during optimization, is further emphasized by the process of flattening 2.5D and 3D networks, as detailed in \cref{sec:flattening}. 

Some residual errors may remain after optimizing the vertices. To resolve these residual errors, the network is scaled down with a scalar 
$s= \min\left(\textbf{l}^1/\textbf{l}^0\right)$, such that all edge lengths in the network are shorter than or equal to their unstretched counterparts ($s\textbf{l}^1\leq\textbf{l}^0$).  

The final step is to turn each edge into an arc with an arc length $l^0$. Each arc is defined by a radius $R$ and an angle $\alpha$, which can be determined for a generic edge by solving
\begin{align}
    l^0 &= R \alpha  && \text{(Arc length equation)} \label{eq:arc_length} \\
    \frac{s l^1}{2} &= R \sin\left(\frac{\alpha}{2}\right)  && \text{(Trigonometric relation)}  \label{eq:trig}
\end{align}

Rewriting \cref{eq:trig} and \cref{eq:arc_length} yields an expression for the length ratio as a function of the arc angle $\alpha$:
\begin{equation}\label{eq:l_ratio_to_theta}
    \frac{s l^1}{l^0} = \frac{2}{\alpha} \sin\left(\frac{\alpha}{2}\right).
\end{equation}

A cubic interpolation function of \cref{eq:l_ratio_to_theta} is set up to avoid solving a transcendental equation every time an arc angle needs to be determined from a length ratio. Further details can be found in~\ref{Ap:arc_params}. One final post-processing step is considered: leaf edges. Consider an edge of which one vertex is connected to only one other vertex. After scaling down the network, the leaf edges can be made to the exact desired length without penalty. Leaf edges are accounted for by moving their free vertex in the direction of the edge's long axis such that the edge has a length of $l^0$.

\begin{figure}
    \centering
    \begin{subfigure}{0.45\textwidth}
        \centering
        \begin{tikzpicture}[scale=0.35]
            \coordinate (A) at (-10.        ,  10.        );
            \coordinate (B) at ( -5,   5);
            \coordinate (C) at (  0.,   0.);
            \coordinate (D) at (  5,  -5);
            \coordinate (E) at ( 10.        , -10.        );
            \coordinate (F) at (-10.        , -10.        );
            \coordinate (G) at ( -5,  -5);
            \coordinate (H) at (  5,   5);
            \coordinate (I) at ( 10.        ,  10.        );
                    
           \draw (A) -- (B) node[midway, right] {\small 11.1};
            \draw (B) -- (C) node[midway, right] {\small 4.03};
            \draw (C) -- (D) node[midway, right] {\small 4.03};
            \draw (D) -- (E) node[midway, right] {\small 13.1};
            \draw (F) -- (G) node[midway, right] {\small 12.1};
            \draw (G) -- (C) node[midway, right] {\small 3.02};
            \draw (C) -- (H) node[midway, right] {\small 5.03};
            \draw (H) -- (I) node[midway, right] {\small 12.1};
            \draw (B) -- (G) node[midway, left] {\small 3.52};
            \draw (G) -- (D) node[midway, above] {\small 4.53};
            \draw (D) -- (H) node[midway, right] {\small 4.03};
            \draw (H) -- (B) node[midway, above] {\small 4.03};

            \foreach \point in {A,E,F,I}
                \filldraw (\point) circle (5pt);
        \end{tikzpicture}
        \caption*{(a)}
        \label{fig:unit-figure-topology}
    \end{subfigure}
    \hspace{0.5em}
    \raisebox{4cm}{\LARGE$\mathbf{\rightarrow}$}
    \hspace{0.5em}
    \begin{subfigure}{0.45\textwidth}
        \centering
         \begin{tikzpicture}[scale=0.35]
            \coordinate (A) at (-10.        ,  10.        );
            \coordinate (B) at ( -4.5877918 ,   4.98509513);
            \coordinate (C) at (  0.82917619,   0.43813642);
            \coordinate (D) at (  5.17568385,  -5.22586088);
            \coordinate (E) at ( 10.        , -10.        );
            \coordinate (F) at (-10.        , -10.        );
            \coordinate (G) at ( -4.79474632,  -5.42409268);
            \coordinate (H) at (  5.05989797,   4.84910477);
            \coordinate (I) at ( 10.        ,  10.        );

            \draw[thick, color = {rgb,1:red,0.710; green,0.869;blue, 0.169}] (A) -- (B);
            \draw[thick, color = {rgb,1:red,0.282; green,0.090;blue, 0.412}] (B) -- (C);
            \draw[thick, color = {rgb,1:red,0.282; green,0.095;blue, 0.417}] (C) -- (D);
            \draw[thick, color = {rgb,1:red,0.993; green,0.906;blue, 0.144}] (D) -- (E);
            \draw[thick, color = {rgb,1:red,0.794; green,0.881;blue, 0.120}] (F) -- (G);
            \draw[thick, color = {rgb,1:red,0.267; green,0.005;blue, 0.329}] (G) -- (C);
            \draw[thick, color = {rgb,1:red,0.283; green,0.136;blue, 0.453}] (C) -- (H);
            \draw[thick, color = {rgb,1:red,0.896; green,0.894;blue, 0.096}] (H) -- (I);
            \draw[thick, color = {rgb,1:red,0.259; green,0.252;blue, 0.525}] (B) -- (G);
            \draw[thick, color = {rgb,1:red,0.196; green,0.395;blue, 0.555}] (G) -- (D);
            \draw[thick, color = {rgb,1:red,0.232; green,0.318;blue, 0.545}] (D) -- (H);
            \draw[thick, color = {rgb,1:red,0.243; green,0.292;blue, 0.539}] (H) -- (B);
            
            \foreach \point in {A,E,F,I}
                \filldraw (\point) circle (5pt);

                \begin{axis}[
                    hide axis,
                    scale only axis,
                    height=10cm,
                    width=11cm,
                    colormap/viridis,
                    colorbar,
                    point meta min=0.18,
                    point meta max=0.65,
                    colorbar style={
                        height=20cm,
                        font=\Large,
                        ylabel style={font=\Large},
                        tick label style={font=\Large},
                        ytick={0.18, 0.3, 0.4, 0.50, 0.61} 
                    }
                ]
                \end{axis}
        \end{tikzpicture}
        \caption*{(b)\LARGE$\downarrow$}
        \label{fig:unit-figure-forces}
    \end{subfigure}
    \begin{subfigure}{0.45\textwidth}
        \centering
        \begin{tikzpicture}[scale=0.35]
        \draw (-8.089, 8.217) -- (-4.194, 4.568);
        \draw (-4.194, 4.568) arc (211.059:250.791:9.521);
        \draw (0.830, 0.489) arc (196.202:235.517:9.707);
        \draw (4.655, -4.804) -- (7.942, -7.975);
        \draw (-8.018, -8.260) -- (-4.282, -5.051);
       \draw (-4.282, -5.051) arc (336.089:298.521:-11.705);
        \draw (0.830, 0.489) arc (294.531:336.507:7.708);
        \draw (4.698, 4.428) -- (8.111, 8.025);
        \draw (-4.194, 4.568) arc (171.531:187.422:34.796);
        \draw (-4.282, -5.051) arc (263.713:279.458:32.637);
        \draw (4.655, -4.804) -- (4.698, 4.428);
        \draw (4.698, 4.428) arc (87.754:90.448:189.142);

        \draw[dashed] (-4.282, -5.051) arc (298.521:336.089:11.705);

        \coordinate (A0) at (-10.        ,  10.        );
        \coordinate (E0) at ( 10.        , -10.        );
        \coordinate (F0) at (-10.        , -10.        );
        \coordinate (I0) at ( 10.        ,  10.        );   
        \foreach \point in {A0,E0,F0,I0}
                \filldraw (\point) circle (5pt);
        \end{tikzpicture}
        \caption*{(d)}
        \label{fig:unit-figure-arcs}
    \end{subfigure}
        \hspace{0.5em}
    \raisebox{4cm}{\LARGE$\mathbf{\leftarrow}$}
    \hspace{0.5em}
    \begin{subfigure}{0.45\textwidth}
        \centering
        \begin{tikzpicture}[scale=0.35]
            \coordinate (A) at (-8.08854089,  8.2168042 );
            \coordinate (B) at (-4.19378708,  4.56792606);
            \coordinate (C) at ( 0.82980972,  0.48897919);
            \coordinate (D) at ( 4.65545589, -4.80384375);
            \coordinate (E) at ( 7.94169568, -7.97488877);
            \coordinate (F) at (-8.01807825, -8.26002251);
            \coordinate (G) at (-4.28168326, -5.05128022);
            \coordinate (H) at ( 4.69849867,  4.42841963);
            \coordinate (I) at ( 8.1107842 ,  8.02478071);
        
            \coordinate (A0) at (-10.        ,  10.        );
            \coordinate (E0) at ( 10.        , -10.        );
            \coordinate (F0) at (-10.        , -10.        );
            \coordinate (I0) at ( 10.        ,  10.        );            
            
            \draw[thick, red] (A) -- (B);
            \draw[thick, red] (B) -- (C);
            \draw[thick, red] (C) -- (D);
            \draw[thick, red] (D) -- (E);
            \draw[thick, red] (F) -- (G);
            \draw[thick, red] (G) -- (C);
            \draw[thick, red] (C) -- (H);
            \draw[thick, red] (H) -- (I);
            \draw[thick, red] (B) -- (G);
            \draw[thick, red] (G) -- (D);
            \draw[thick, red] (D) -- (H);
            \draw[thick, red] (H) -- (B);

            \foreach \point in {A0,E0,F0,I0}
                \filldraw (\point) circle (5pt);
        \end{tikzpicture}
        \caption*{(c)}
        \label{fig:unit-figure-optimized}
    \end{subfigure}
    \caption{a) A unit cell topology with user-described force densities $q$ times a thousand (N/mm) labeled at each edge. The corner circles indicate fixed points $\textbf{x}_f$. b) The stretched shape as found with the force density method, see \cref{sec:FDM}. The color represents the tension in the edge in Newton. c) The shape of the network after optimization and scaling. d) Each edge is converted to an arc with arc length $l^0$. Note the dashed arc: Edges can be turned into arcs in two directions. Directions should be flipped to ensure no overlapping features.}
    \label{fig:design-validation-method}
\end{figure}
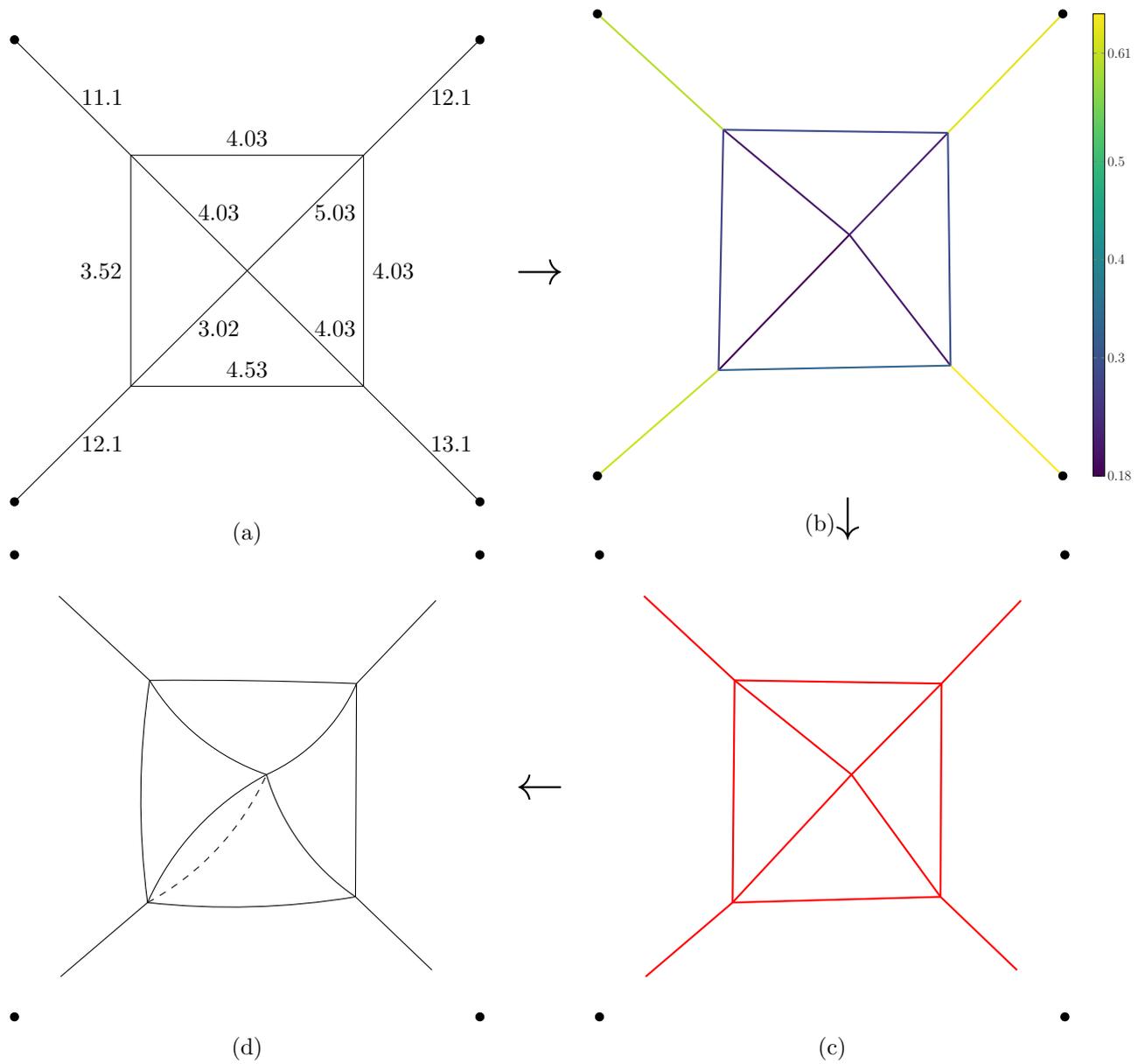

\subsection{Network fabrication} \label{sec:fabrication}
The steps to convert the flattened, relaxed, and scaled networks into machine code compatible with FFF are the following. Decompose the network into continuous paths (\cref{sec:printable paths}). Account  for intersections between the printed path and previously printed paths (\cref{sec:intersections}).

\subsubsection{Path decomposition} \label{sec:printable paths}
\begin{figure}
    \centering
    \begin{subfigure}{0.18\textwidth}
        \centering
        \begin{tikzpicture}[scale=0.15]
            \coordinate (A) at (-10, 0);
            \coordinate (B) at (-5, 0);
            \coordinate (C) at (0, 0);
            \coordinate (D) at (5, 0);
            \coordinate (E) at (10, 0);
            \coordinate (F) at (0, -10);
            \coordinate (G) at (0, -5);
            \coordinate (H) at (0, 5);
            \coordinate (I) at (0, 10);
        
            \draw (A) -- (B);
            \draw (B) -- (C);
            \draw (C) -- (D);
            \draw (D) -- (E);
            \draw (F) -- (G);
            \draw (G) -- (C);
            \draw (C) -- (H);
            \draw (H) -- (I);
            \draw (B) -- (G);
            \draw (G) -- (D);
            \draw (D) -- (H);
            \draw (H) -- (B);

            \foreach \point in {A,B,C,D,E,F,G,H,I}
                \filldraw (\point) circle (3pt);
            \foreach \point in {A,E,F,I}
                \filldraw (\point) circle (5pt);
        \end{tikzpicture}
        \caption{}
        \label{fig:unit-cell}
    \end{subfigure}
    \hspace{0.5em}
    \raisebox{1.5cm}{$\mathbf{=}$}
    \hspace{0.5em}
    \begin{subfigure}{0.18\textwidth}
        \centering
        \begin{tikzpicture}[scale=0.15]
            \coordinate (A) at (-10, 0);
            \coordinate (B) at (-5, 0);
            \coordinate (C) at (0, 0);
            \coordinate (D) at (5, 0);
            \coordinate (E) at (10, 0);
            \coordinate (F) at (0, -10);
            \coordinate (G) at (0, -5);
            \coordinate (H) at (0, 5);
            \coordinate (I) at (0, 10);
        
            \draw (A) -- (B);
            \draw (B) -- (C);
            \draw (C) -- (D);
            \draw (D) -- (E);
            \foreach \point in {A,B,C,D,E,F,G,H,I}
                \filldraw (\point) circle (3pt);
            \foreach \point in {A,E,F,I}
                \filldraw (\point) circle (5pt);
        \end{tikzpicture}
        \caption{}
    \end{subfigure}
    \hspace{0.5em}
    \raisebox{1.5cm}{$\mathbf{+}$}
    \hspace{0.5em}
    \begin{subfigure}{0.18\textwidth}
        \centering
        \begin{tikzpicture}[scale=0.15]
            \coordinate (A) at (-10, 0);
            \coordinate (B) at (-5, 0);
            \coordinate (C) at (0, 0);
            \coordinate (D) at (5, 0);
            \coordinate (E) at (10, 0);
            \coordinate (F) at (0, -10);
            \coordinate (G) at (0, -5);
            \coordinate (H) at (0, 5);
            \coordinate (I) at (0, 10);
        
            \draw (F) -- (G);
            \draw (G) -- (C);
            \draw (C) -- (H);
            \draw (H) -- (I);
            \foreach \point in {A,B,C,D,E,F,G,H,I}
                \filldraw (\point) circle (3pt);
            \foreach \point in {A,E,F,I}
                \filldraw (\point) circle (5pt);
        \end{tikzpicture}
        \caption{}
    \end{subfigure}
    \hspace{0.5em}
    \raisebox{1.5cm}{$\mathbf{+}$}
    \hspace{0.5em}
    \begin{subfigure}{0.18\textwidth}
        \centering
        \begin{tikzpicture}[scale=0.15]
            \coordinate (A) at (-10, 0);
            \coordinate (B) at (-5, 0);
            \coordinate (C) at (0, 0);
            \coordinate (D) at (5, 0);
            \coordinate (E) at (10, 0);
            \coordinate (F) at (0, -10);
            \coordinate (G) at (0, -5);
            \coordinate (H) at (0, 5);
            \coordinate (I) at (0, 10);
        
            \draw (B) -- (G);
            \draw (G) -- (D);
            \draw (D) -- (H);
            \draw (H) -- (B);

            \foreach \point in {A,B,C,D,E,F,G,H,I}
                \filldraw (\point) circle (3pt);
            \foreach \point in {A,E,F,I}
                \filldraw (\point) circle (5pt);
        \end{tikzpicture}
        \caption{}
    \end{subfigure}
    \caption{A unit cell network (a) is used for validating the method. The unit cell can be decomposed into parts b, c, and d.}
    \label{fig: unit-cell}
\end{figure}
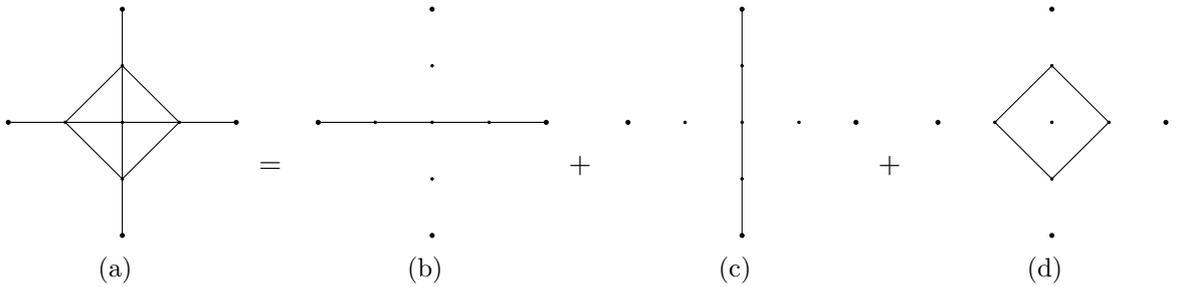

The FFF process is most unstable at the beginning and end of a print path, primarily due to the viscoelastic behavior of the melted polymers. This viscoelasticity introduces a delay between the intended and actual start/stop locations, which can compromise printing precision. Therefore, minimizing the number of start-stop events by printing structures in as few continuous paths as possible is highly desirable. In this work, structures are manually decomposed into printable paths. For example, a unit cell structure, later used for validation, can be divided into three sections: a horizontal path, a vertical path, and a loop, as shown in \cref{fig:unit-cell}.

Automatic decomposition of networks into continuous edge-covering paths is possible using optimization-based methods~\cite{Botler2019}. However, computing exact solutions is an NP-hard problem, which limits the feasibility for large graphs. To address this, developments such as the Hybrid Lagrangian Relaxation and Particle Swarm Optimization (LaPSO) approach offer relatively scalable approximations that yield practical solutions for complex networks~\cite{Weiner2021}. These methods provide a promising foundation for automating and generalizing the decomposition process, especially when manual design becomes impractical.

\subsubsection{Intersections} \label{sec:intersections}
After decomposing the network into printable paths, intersections between these paths must be identified. For each printable path, a set of intersecting vertices is computed by finding the intersection between its vertices and the union of the vertices from all previously printed paths. When the print path approaches an intersection, the print nozzle raises to avoid collision. This raising follows a linear trajectory, beginning 0.6 mm before the intersection point (1.5 times the nozzle diameter) and lifting by one layer height. The transition is designed with a slight overlap to ensure good adhesion between layers. A parameter study confirmed that this approach results in reliable adhesion with minimal damage to previously printed paths.
\subsection{Flattening and accounting for crossing edges} \label{sec:flattening}
To print curved (2.5D) or 3D networks on a flat 2D printer bed, networks must first be flattened, i.e., projected onto a planar surface. It is vital that the original geometric layout of the network is preserved as much as possible to preserve reliable manufacturability.
The proposed flattening procedure is a coordinate transformation from Cartesian space to a polar or spherical coordinate system defined with respect to a user-defined unit vector $\textbf{t}$ and center point $\textbf{c}$. The final flattening procedure consists of disregarding vertex-specific radii and using a mean radius for all vertices. The procedure is set out in more detail in~\ref{Ap:flattening_methods}. 

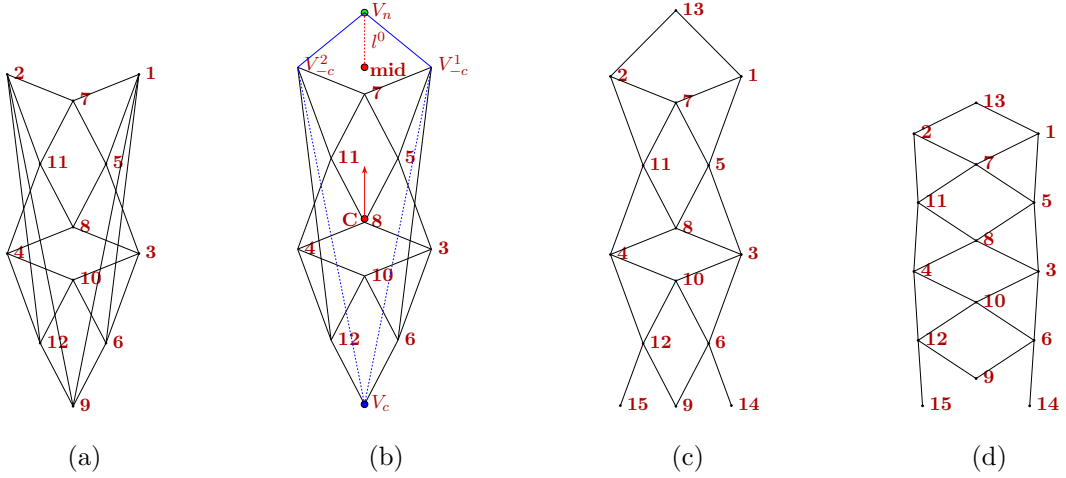
\begin{figure}
    \centering
    \begin{subfigure}{0.23\textwidth}
        \centering
        \rotatebox{90}{ 
            \resizebox{1.3\textwidth}{!}{%
                \begin{tikzpicture}
\draw[fill=black] (10.966, -4.000) circle (0.05);
\node[right, rotate=-90, text=red!70!black, scale=3] at (11.066, -4.000) {\textbf{1}};
\draw[fill=black] (10.966, 4.000) circle (0.05);
\node[right, rotate=-90, text=red!70!black, scale=3] at (11.066, 4.000) {\textbf{2}};
\draw[fill=black] (0.000, -4.000) circle (0.05);
\node[right, rotate=-90, text=red!70!black, scale=3] at (0.100, -4.000) {\textbf{3}};
\draw[fill=black] (0.000, 4.000) circle (0.05);
\node[right, rotate=-90, text=red!70!black, scale=3] at (0.100, 4.000) {\textbf{4}};
\draw[fill=black] (5.483, -2.000) circle (0.05);
\node[right, rotate=-90, text=red!70!black, scale=3] at (5.583, -2.000) {\textbf{5}};
\draw[fill=black] (-5.483, -2.000) circle (0.05);
\node[right, rotate=-90, text=red!70!black, scale=3] at (-5.383, -2.000) {\textbf{6}};
\draw[fill=black] (9.348, 0.000) circle (0.05);
\node[right, rotate=-90, text=red!70!black, scale=3] at (9.448, 0.000) {\textbf{7}};
\draw[fill=black] (1.618, 0.000) circle (0.05);
\node[right, rotate=-90, text=red!70!black, scale=3] at (1.718, 0.000) {\textbf{8}};
\draw[fill=black] (-9.348, 0.000) circle (0.05);
\node[right, rotate=-90, text=red!70!black, scale=3] at (-9.248, 0.000) {\textbf{9}};
\draw[fill=black] (-1.618, 0.000) circle (0.05);
\node[right, rotate=-90, text=red!70!black, scale=3] at (-1.518, 0.000) {\textbf{10}};
\draw[fill=black] (5.483, 2.000) circle (0.05);
\node[right, rotate=-90, text=red!70!black, scale=3] at (5.583, 2.000) {\textbf{11}};
\draw[fill=black] (-5.483, 2.000) circle (0.05);
\node[right, rotate=-90, text=red!70!black, scale=3] at (-5.383, 2.000) {\textbf{12}};
\draw[black] (9.348, 0.000) -- (10.966, -4.000);
\draw[black] (10.966, -4.000) -- (-9.348, 0.000);
\draw[black] (10.966, -4.000) -- (5.483, -2.000);
\draw[black] (10.966, -4.000) -- (-5.483, -2.000);
\draw[black] (0.000, -4.000) -- (5.483, -2.000);
\draw[black] (0.000, -4.000) -- (-5.483, -2.000);
\draw[black] (0.000, -4.000) -- (1.618, 0.000);
\draw[black] (0.000, -4.000) -- (-1.618, 0.000);
\draw[black] (5.483, -2.000) -- (9.348, 0.000);
\draw[black] (5.483, -2.000) -- (1.618, 0.000);
\draw[black] (-5.483, -2.000) -- (-9.348, 0.000);
\draw[black] (-5.483, -2.000) -- (-1.618, 0.000);
\draw[black] (10.966, 4.000) -- (9.348, 0.000);
\draw[black] (9.348, 0.000) -- (5.483, 2.000);
\draw[black] (10.966, 4.000) -- (-9.348, 0.000);
\draw[black] (-9.348, 0.000) -- (-5.483, 2.000);
\draw[black] (1.618, 0.000) -- (5.483, 2.000);
\draw[black] (0.000, 4.000) -- (1.618, 0.000);
\draw[black] (-1.618, 0.000) -- (-5.483, 2.000);
\draw[black] (0.000, 4.000) -- (-1.618, 0.000);
\draw[black] (10.966, 4.000) -- (5.483, 2.000);
\draw[black] (0.000, 4.000) -- (5.483, 2.000);
\draw[black] (10.966, 4.000) -- (-5.483, 2.000);
\draw[black] (0.000, 4.000) -- (-5.483, 2.000);
\end{tikzpicture}
            }
        }
        \caption{}
        \label{fig:unwrapped_tensegrity0}
    \end{subfigure}
    \hspace{0.002\textwidth}
    \begin{subfigure}{0.23\textwidth}
        \centering
        \rotatebox{90}{ 
            \resizebox{1.53\textwidth}{!}{%
                \begin{tikzpicture}
\coordinate (V1) at (10.966, -4.000);
\coordinate (V2) at (10.966, 4.000);
\coordinate (V3) at (0.000, -4.000);
\coordinate (V4) at (0.000, 4.000);
\coordinate (V5) at (5.483, -2.000);
\coordinate (V6) at (-5.483, -2.000);
\coordinate (V7) at (9.348, 0.000);
\coordinate (V8) at (1.618, 0.000);
\coordinate (V9) at (-9.348, 0.000);
\coordinate (V10) at (-1.618, 0.000);
\coordinate (V11) at (5.483, 2.000);
\coordinate (V12) at (-5.483, 2.000);

\node[right, rotate=-90, text=red!70!black, scale=3] at ($(V1) + (0.1,0)$) {$V_{-c}^1$};
\node[right, rotate=-90, text=red!70!black, scale=3] at ($(V2) + (0.1,0)$) {$V_{-c}^2$};
\node[right, rotate=-90, text=red!70!black, scale=3] at ($(V3) + (0.1,0)$) {\textbf{3}};
\node[right, rotate=-90, text=red!70!black, scale=3] at ($(V4) + (0.1,0)$) {\textbf{4}};
\node[right, rotate=-90, text=red!70!black, scale=3] at ($(V5) + (0.1,0)$) {\textbf{5}};
\node[right, rotate=-90, text=red!70!black, scale=3] at ($(V6) + (0.1,0)$) {\textbf{6}};
\node[right, rotate=-90, text=red!70!black, scale=3] at ($(V7) + (0.1,0)$) {\textbf{7}};
\node[right, rotate=-90, text=red!70!black, scale=3] at ($(V8) + (0.1,0)$) {\textbf{8}};
\node[right, rotate=-90, text=red!70!black, scale=3] at ($(V9) + (0.1,0)$) {$V_c$};
\node[right, rotate=-90, text=red!70!black, scale=3] at ($(V10) + (0.1,0)$) {\textbf{10}};
\node[right, rotate=-90, text=red!70!black, scale=3] at ($(V11) + (0.1,0)$) {\textbf{11}};
\node[right, rotate=-90, text=red!70!black, scale=3] at ($(V12) + (0.1,0)$) {\textbf{12}};

\def\edges{{7/1}, {1/5}, {1/6}, {3/5}, {3/6}, {3/8}, {3/10},
           {5/7}, {5/8}, {6/9}, {6/10}, {2/7}, {7/11}, {9/12},
           {8/11}, {4/8}, {10/12}, {4/10}, {2/11}, {4/11}, {2/12}, {4/12}}

\foreach \a/\b in \edges {
    \draw[black] (V\a) -- (V\b);
}

\coordinate (V13) at (1.828, 0.); 
\coordinate (V14) at (10.966, 0.); 
\coordinate (V15) at (14.257, 0.); 
\coordinate (V16) at (5, 0.); 
\coordinate (V17) at (12.5, 0.); 
\draw[fill=red] (V13) circle (0.2);

\draw[fill=blue] (V9) circle (0.2);
\draw[fill=red] (V14) circle (0.2);
\draw[fill=green] (V15) circle (0.2);

\draw[blue, dashed] (V9) -- (V1);
\draw[blue, dashed] (V9) -- (V2);

\draw[-{Stealth[length=12pt,width=8pt]}, very thick, red] (V13) -- (V16);
\draw[red, dashed] (V15) -- (V14);

\draw[blue, thick] (V15) -- (V1);
\draw[blue, thick] (V15) -- (V2);

\node[right, rotate=-90, text=red!70!black, scale=3] at ($(V15) + (0.1,0)$) {$V_n$};

\node[right, rotate=-90, text=red!70!black, scale=3] at ($(V17) + (0.1,0)$) {$l^0$};

\node[left, rotate=-90, text=red!70!black, scale=3] at ($(V8) + (0.2,0.)$) {\textbf{C}};
\node[right, rotate=-90, text=red!70!black, scale=3] at ($(V14) + (-0.1,0.1)$) {\textbf{mid}};

\end{tikzpicture}
            }
        }
        \caption{}
        \label{fig:unwrapped_tensegrity0.5}
    \end{subfigure}
    \hspace{0.002\textwidth}
    \begin{subfigure}{0.23\textwidth}
        \centering
        \rotatebox{90}{ 
            \resizebox{1.53\textwidth}{!}{%
                \begin{tikzpicture}
\draw[fill=black] (10.966, -4.000) circle (0.05);
\node[right, rotate=-90, text=red!70!black, scale=3] at (11.066, -4.000) {\textbf{1}};
\draw[fill=black] (10.966, 4.000) circle (0.05);
\node[right, rotate=-90, text=red!70!black, scale=3] at (11.066, 4.000) {\textbf{2}};
\draw[fill=black] (0.000, -4.000) circle (0.05);
\node[right, rotate=-90, text=red!70!black, scale=3] at (0.100, -4.000) {\textbf{3}};
\draw[fill=black] (0.000, 4.000) circle (0.05);
\node[right, rotate=-90, text=red!70!black, scale=3] at (0.100, 4.000) {\textbf{4}};
\draw[fill=black] (5.483, -2.000) circle (0.05);
\node[right, rotate=-90, text=red!70!black, scale=3] at (5.583, -2.000) {\textbf{5}};
\draw[fill=black] (-5.483, -2.000) circle (0.05);
\node[right, rotate=-90, text=red!70!black, scale=3] at (-5.383, -2.000) {\textbf{6}};
\draw[fill=black] (9.348, 0.000) circle (0.05);
\node[right, rotate=-90, text=red!70!black, scale=3] at (9.448, 0.000) {\textbf{7}};
\draw[fill=black] (1.618, 0.000) circle (0.05);
\node[right, rotate=-90, text=red!70!black, scale=3] at (1.718, 0.000) {\textbf{8}};
\draw[fill=black] (-9.348, 0.000) circle (0.05);
\node[right, rotate=-90, text=red!70!black, scale=3] at (-9.248, 0.000) {\textbf{9}};
\draw[fill=black] (-1.618, 0.000) circle (0.05);
\node[right, rotate=-90, text=red!70!black, scale=3] at (-1.518, 0.000) {\textbf{10}};
\draw[fill=black] (5.483, 2.000) circle (0.05);
\node[right, rotate=-90, text=red!70!black, scale=3] at (5.583, 2.000) {\textbf{11}};
\draw[fill=black] (-5.483, 2.000) circle (0.05);
\node[right, rotate=-90, text=red!70!black, scale=3] at (-5.383, 2.000) {\textbf{12}};
\draw[fill=black] (15.032, 0.000) circle (0.05);
\node[right, rotate=-90, text=red!70!black, scale=3] at (15.132, 0.000) {\textbf{13}};
\draw[fill=black] (-9.303, -3.393) circle (0.05);
\node[right, rotate=-90, text=red!70!black, scale=3] at (-9.203, -3.393) {\textbf{14}};
\draw[fill=black] (-9.303, 3.393) circle (0.05);
\node[right, rotate=-90, text=red!70!black, scale=3] at (-9.203, 3.393) {\textbf{15}};
\draw[black] (9.348, 0.000) -- (10.966, -4.000);
\draw[black] (10.966, -4.000) -- (15.032, 0.000);
\draw[black] (10.966, -4.000) -- (5.483, -2.000);
\draw[black] (-9.303, -3.393) -- (-5.483, -2.000);
\draw[black] (0.000, -4.000) -- (5.483, -2.000);
\draw[black] (0.000, -4.000) -- (-5.483, -2.000);
\draw[black] (0.000, -4.000) -- (1.618, 0.000);
\draw[black] (0.000, -4.000) -- (-1.618, 0.000);
\draw[black] (5.483, -2.000) -- (9.348, 0.000);
\draw[black] (5.483, -2.000) -- (1.618, 0.000);
\draw[black] (-5.483, -2.000) -- (-9.348, 0.000);
\draw[black] (-5.483, -2.000) -- (-1.618, 0.000);
\draw[black] (10.966, 4.000) -- (9.348, 0.000);
\draw[black] (9.348, 0.000) -- (5.483, 2.000);
\draw[black] (10.966, 4.000) -- (15.032, 0.000);
\draw[black] (-9.348, 0.000) -- (-5.483, 2.000);
\draw[black] (1.618, 0.000) -- (5.483, 2.000);
\draw[black] (0.000, 4.000) -- (1.618, 0.000);
\draw[black] (-1.618, 0.000) -- (-5.483, 2.000);
\draw[black] (0.000, 4.000) -- (-1.618, 0.000);
\draw[black] (10.966, 4.000) -- (5.483, 2.000);
\draw[black] (0.000, 4.000) -- (5.483, 2.000);
\draw[black] (-9.303, 3.393) -- (-5.483, 2.000);
\draw[black] (0.000, 4.000) -- (-5.483, 2.000);
\end{tikzpicture}
            }
        }
        \caption{}
        \label{fig:unwrapped_tensegrity1}
    \end{subfigure}
    \hspace{0.002\textwidth}
    \begin{subfigure}{0.23\textwidth}
        \centering
        \rotatebox{90}{ 
            \resizebox{1.2\textwidth}{!}{%
                \begin{tikzpicture}
\draw[fill=black] (8.652, -3.638) circle (0.05);
\node[right, rotate=-90, text=red!70!black, scale=3] at (8.752, -3.638) {\textbf{1}};
\draw[fill=black] (8.651, 3.638) circle (0.05);
\node[right, rotate=-90, text=red!70!black, scale=3] at (8.751, 3.638) {\textbf{2}};
\draw[fill=black] (0.535, -3.640) circle (0.05);
\node[right, rotate=-90, text=red!70!black, scale=3] at (0.635, -3.640) {\textbf{3}};
\draw[fill=black] (0.535, 3.640) circle (0.05);
\node[right, rotate=-90, text=red!70!black, scale=3] at (0.635, 3.640) {\textbf{4}};
\draw[fill=black] (4.593, -3.390) circle (0.05);
\node[right, rotate=-90, text=red!70!black, scale=3] at (4.693, -3.390) {\textbf{5}};
\draw[fill=black] (-3.523, -3.388) circle (0.05);
\node[right, rotate=-90, text=red!70!black, scale=3] at (-3.423, -3.388) {\textbf{6}};
\draw[fill=black] (6.837, 0.000) circle (0.05);
\node[right, rotate=-90, text=red!70!black, scale=3] at (6.937, 0.000) {\textbf{7}};
\draw[fill=black] (2.348, -0.000) circle (0.05);
\node[right, rotate=-90, text=red!70!black, scale=3] at (2.448, -0.000) {\textbf{8}};
\draw[fill=black] (-5.771, -0.000) circle (0.05);
\node[right, rotate=-90, text=red!70!black, scale=3] at (-5.671, -0.000) {\textbf{9}};
\draw[fill=black] (-1.276, -0.000) circle (0.05);
\node[right, rotate=-90, text=red!70!black, scale=3] at (-1.176, -0.000) {\textbf{10}};
\draw[fill=black] (4.593, 3.390) circle (0.05);
\node[right, rotate=-90, text=red!70!black, scale=3] at (4.693, 3.390) {\textbf{11}};
\draw[fill=black] (-3.523, 3.388) circle (0.05);
\node[right, rotate=-90, text=red!70!black, scale=3] at (-3.423, 3.388) {\textbf{12}};
\draw[fill=black] (10.467, 0.000) circle (0.05);
\node[right, rotate=-90, text=red!70!black, scale=3] at (10.567, 0.000) {\textbf{13}};
\draw[fill=black] (-7.380, -3.131) circle (0.05);
\node[right, rotate=-90, text=red!70!black, scale=3] at (-7.280, -3.131) {\textbf{14}};
\draw[fill=black] (-7.380, 3.131) circle (0.05);
\node[right, rotate=-90, text=red!70!black, scale=3] at (-7.280, 3.131) {\textbf{15}};
\draw[black] (6.837, 0.000) -- (8.652, -3.638);
\draw[black] (8.652, -3.638) -- (10.467, 0.000);
\draw[black] (8.652, -3.638) -- (4.593, -3.390);
\draw[black] (-7.380, -3.131) -- (-3.523, -3.388);
\draw[black] (0.535, -3.640) -- (4.593, -3.390);
\draw[black] (0.535, -3.640) -- (-3.523, -3.388);
\draw[black] (0.535, -3.640) -- (2.348, -0.000);
\draw[black] (0.535, -3.640) -- (-1.276, -0.000);
\draw[black] (4.593, -3.390) -- (6.837, 0.000);
\draw[black] (4.593, -3.390) -- (2.348, -0.000);
\draw[black] (-3.523, -3.388) -- (-5.771, -0.000);
\draw[black] (-3.523, -3.388) -- (-1.276, -0.000);
\draw[black] (8.651, 3.638) -- (6.837, 0.000);
\draw[black] (6.837, 0.000) -- (4.593, 3.390);
\draw[black] (8.651, 3.638) -- (10.467, 0.000);
\draw[black] (-5.771, -0.000) -- (-3.523, 3.388);
\draw[black] (2.348, -0.000) -- (4.593, 3.390);
\draw[black] (0.535, 3.640) -- (2.348, -0.000);
\draw[black] (-1.276, -0.000) -- (-3.523, 3.388);
\draw[black] (0.535, 3.640) -- (-1.276, -0.000);
\draw[black] (8.651, 3.638) -- (4.593, 3.390);
\draw[black] (0.535, 3.640) -- (4.593, 3.390);
\draw[black] (-7.380, 3.131) -- (-3.523, 3.388);
\draw[black] (0.535, 3.640) -- (-3.523, 3.388);
\end{tikzpicture}
            }
        }
        \caption{}
        \label{fig:unwrapped_tensegrity2}
    \end{subfigure}
    \caption{Network transformation on the tensegrity network in \cref{sec:tensegrity structure}. a) The cable network of the tensegrity is flattened. $\textbf{c} = (0, 0, 0)$, and no rotations are applied before the Cartesian-to-polar transformation, i.e., $\textbf{t} =[0;0;1]$. b) Steps to remove crossings: Identify the vertex with the most crossings, $V_c = 9$, and duplicate it to create $V_n = 13$. The new vertex $V_n$ is placed along the direction from the network center $C$ toward the midpoint of the complementary vertices $V_{-c} = \{1, 2\}$. Its distance from mid is set to the mean target length $l^0$ of the edges in $E_c$. In the intersecting edges, the old vertex $V_c$ is replaced with $V_n$, shown as dashed blue edges changing to solid blue edges. c) Repeat the process until no crossings are left. d) The network after relaxation.}
    \label{fig:tensegrity_networks}
\end{figure}

The resulting flat 2D network can be seen as the initial guess $\textbf{x}_0$ for the Gauss-Seidel relaxation procedure (\cref{sec:net_opt}). However, the following issue needs to be addressed first. The flattened network can have new internal crossings, thereby altering its inherent topology. The first step to account for these new crossings is to identify all of them. This is achieved by checking all possible edge pairs for crossings, which scales with $O(M^2)$, where $M$ is the number of edges. Despite this quadratic scaling, the process remains efficient for moderately sized networks. For instance, a network with 1,000 edges is expected to process in under 1 second, assuming an edge-edge check takes approximately 0.1 µs.
The resolution of a crossing is explained in \cref{fig:tensegrity_networks}.  
The center of all vertices is defined as $C$. For each vertex involved in crossing edges, the number of crossings it participates in is counted.  The vertex with the most crossings is denoted $V_c = 9$, and its crossing edges are $E_c = [1,9]$ and $[2,9]$. The complementary vertices of these edges are $V_{-c} = \{1, 2\}$. A new vertex $V_n$ is introduced along the direction from $C$ toward the midpoint of $V_{-c}$. Its distance from $C$ is set to the mean target length $l^0$ of the edges in $E_c$. The vertices $V_c$ in edges $E_c$ are then reassigned to $V_n$. This duplication and reassignment process is repeated until all crossings are resolved.

\subsection{Limitations}\label{sec:limitations-methods}
The manufacturability of a network depends strongly on its topology, designed tension gradients, and the extent to which it deviates from a flat geometry, making it challenging to quantify general limitations. In networks with steep tension gradients or in 2.5D/3D configurations where radii vary widely after coordinate transformation, the Gauss-Seidel optimization may cause a vertex to move past a neighbor. This can introduce unintended crossings and alter the network's inherent form. A method to resolve such crossings is discussed in \cref{sec:flattening}, although it introduces additional assembly effort.

Despite the difficulty in quantifying manufacturability in general terms, a theoretical lower bound can be established based on geometric constraints. Take an overdefined network, that is, no vertex position exists that satisfies all desired edge lengths. To overcome this, the methods allow for global scaling and printing of edges as arcs. However, the minimum achievable arc-to-chord length ratio is 0.6366 (see \cref{eq:l_ratio_to_theta} at $\alpha=\pi$), establishing a hard lower bound on manufacturable designs.

The force density method is a form-finding technique for structures composed of elements that bear loads in the axial direction only, such as bars and cables. In this work, the edges of the network are fabricated using flexible polymer TPU and are assumed to behave like cables. This assumption is only valid as long as their lengths are significantly greater than their widths: $l^0\gg w$. It is expected that as networks are printed with shorter edges, the accuracy of the method decreases. This limitation is tested and the results are set out in \cref{sec:limitations-results}.

Similarly, cases with short lengths and large radii need to be considered. In such cases, the length of the edge will vary throughout its thickness. The inner length will be shorter and the outer length longer than the designed length $l^0$.
This limitation is also tested, and the results are presented in \cref{sec:limitations-results}.
\section{Validation}\label{sec:validation}
\begin{figure}[htbp]
    \centering
    \begin{minipage}[t]{0.45\textwidth}
        \centering
        \begin{tikzpicture}
            \node[anchor=south west, inner sep=0pt] (image) at (0,0)
                {\includegraphics[width=\textwidth]{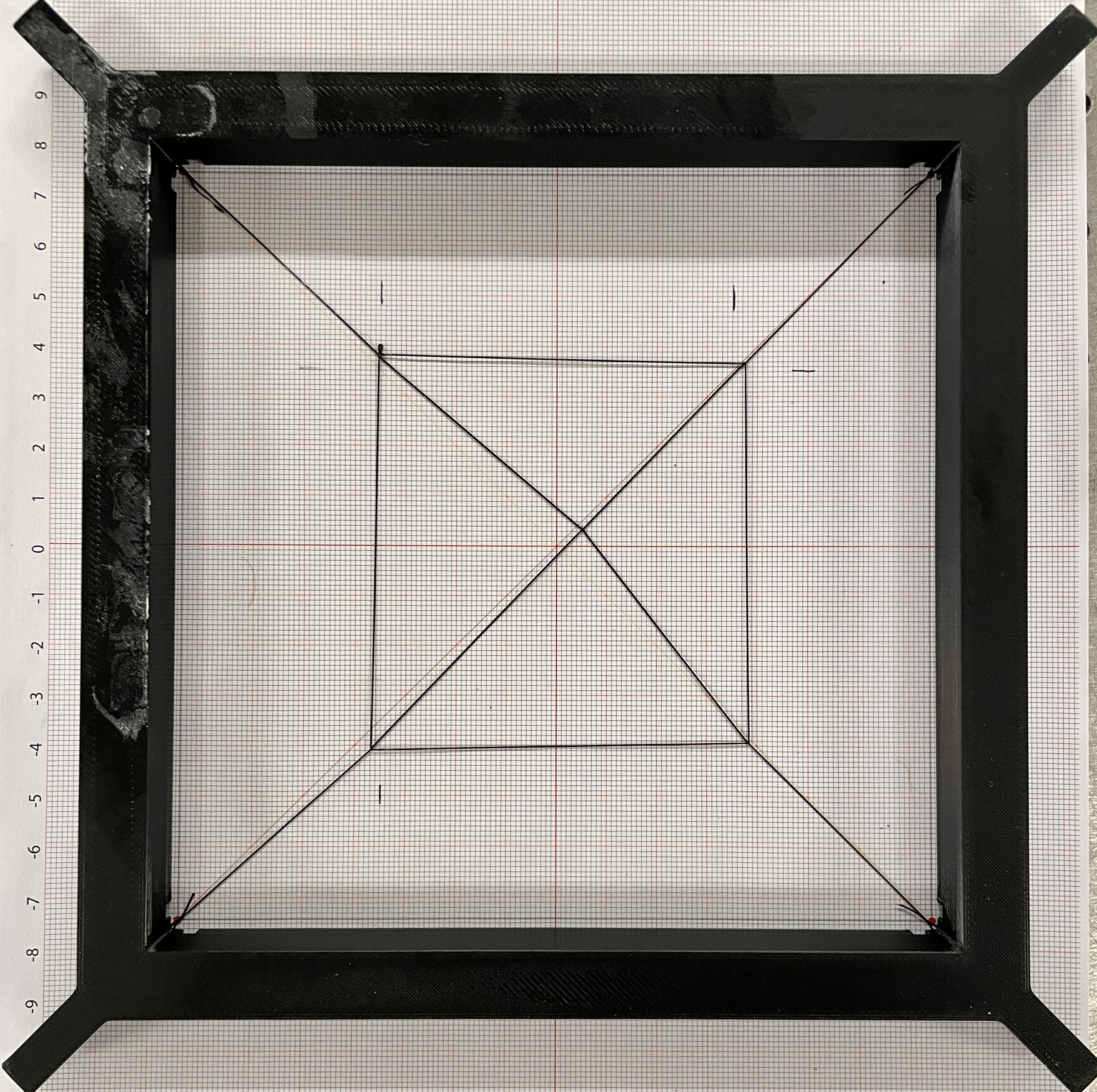}};
            \begin{scope}[x={(image.south east)}, y={(image.north west)}]
                \foreach \i/\x/\y in {
                    1/0.1/0.9,
                    2/0.29/0.7,
                    3/0.48/0.5,
                    4/0.725/0.3,
                    5/0.9/0.1,
                    6/0.1/0.1,
                    7/0.29/0.3,
                    8/0.725/0.7,
                    9/0.9/0.9
                } {
                    \node[draw, fill=white, text=black, thick, font=\large, rounded corners, minimum size=.5cm] at (\x, \y) {\i};
                }
            \end{scope}
        \end{tikzpicture}
    \end{minipage}
    \hspace{1em}
    \raisebox{30mm}{ 
        \begin{minipage}[t]{0.45\textwidth}
            \centering
            \begin{tabular}{|c|c|c|}
    \hline
    \textbf{Vertex No.} & $\textbf{x}$ & $\textbf{x}^m$ \\
    \hline
    1 & (-74.2, 74.2) & - \\
    \hline
    2 & (-34.1, 37.0) & (-34.0, 38.8) \\
    \hline
    3 & (6.16, 3.25) & (6.7, 4.3) \\
    \hline
    4 & (38.4, -38.8) & (39.6, -38.3) \\
    \hline
    5 & (74.2, -74.2) & - \\
    \hline
    6 & (-74.2, -74.2) & - \\
    \hline
    7 & (-35.6, -40.3) & (-34.4, -39.5) \\
    \hline
    8 & (37.6, 36.0) & (38.2, 37.2) \\
    \hline
    9 & (74.2, 74.2) & - \\
    \hline
\end{tabular}        
        \end{minipage}
    }
    \caption{The unit cell corresponding to 7.3 MPa in \cref{fig: error_vs_force} suspended in the frame. The tables shows the designed $\textbf{x}$ and measured $\textbf{x}^m$ vertex locations, enabling evaluating \cref{eq:score} to determine the mean edge length error to be 0.52\%.}
    \label{fig:unit_cell_img}
\end{figure}

The methods for manufacturing networks with programmable tensions are validated on a unit cell structure suspended within a frame. The design process is described in \cref{fig:design-validation-method}, and the printed unit cell is shown in \cref{fig:unit_cell_img}. The unit cell is photographed orthogonally on top of printed paper with a 1 mm–spaced grid, allowing the vertex locations $\textbf{x}^m$ to be measured. Exact vertex positions can be interpolated between the grid lines in the photograph by counting pixels, achieving a measurement tolerance of 0.1 mm. The frame ensures the unit cell is in contact with the grid paper, minimizing image distortion and perspective errors.
The agreement between the measured and designed edge lengths is quantified using 
\begin{equation}
\label{eq:score}
    \text{error} = \frac{1}{M}\sum_{i=1}^M  \frac{|l^{1,m}_i -l^{1}_i|}{l^{1}_i}\cdot100\%
\end{equation}
where the distance is normalized with the target edge lengths to make the error scale-invariant. Since the vertex coordinates and tensions are linked through equilibrium equations (e.g., see \cref{eq:FD-method}), validating the vertex positions implicitly validates the tensions. The mean edge length error of the validation structure in \cref{fig:unit_cell_img} is 0.52\%.

\subsection{Testing limitations}\label{sec:limitations-results}
\begin{figure}[h!]
    \centering
    \begin{subfigure}[t]{.9\textwidth}
    \centering
    \begin{overpic}[width=\textwidth,height=0.5\textheight,keepaspectratio]{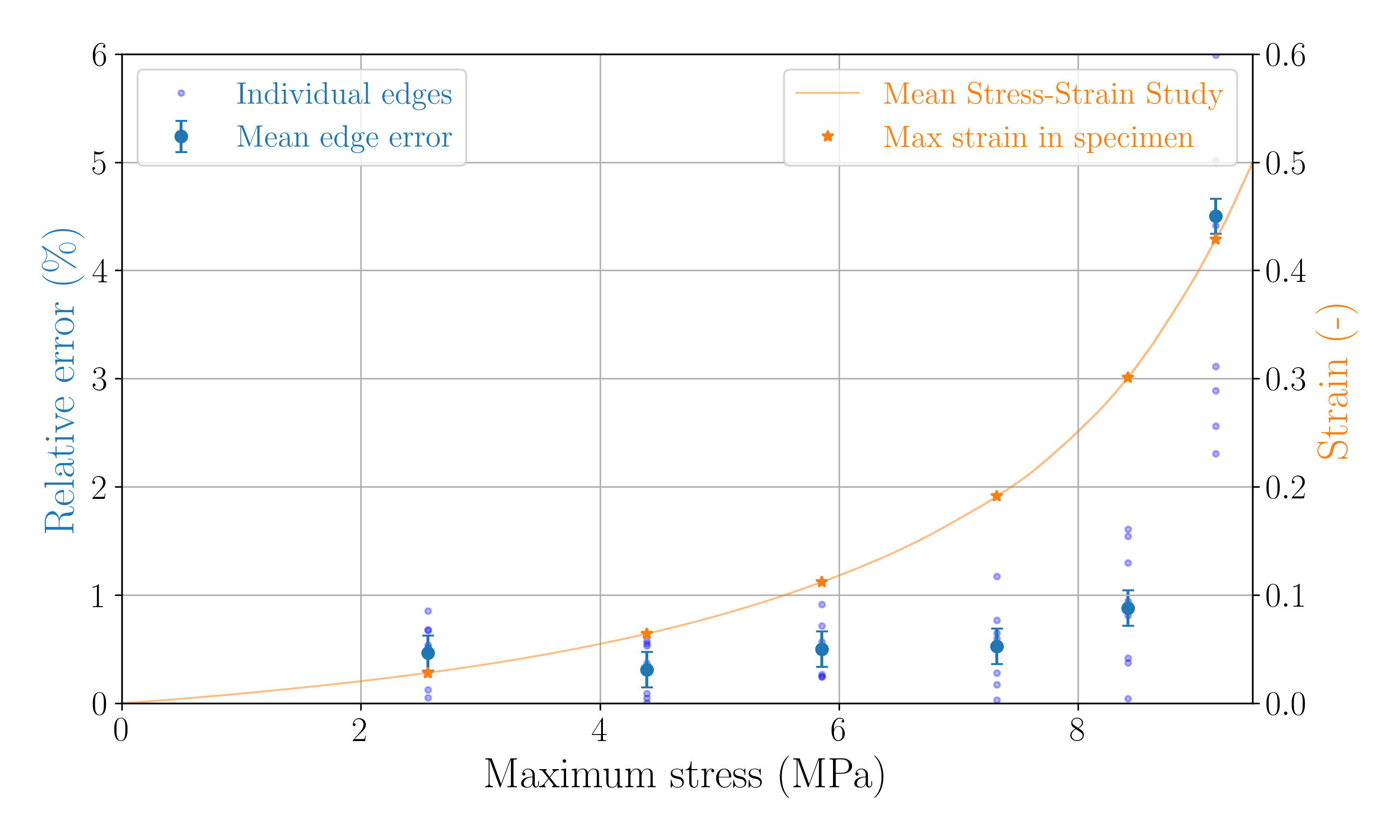}
        \put(15,22){%
            \input{tikzfigures/validation_unit_fstudy_0.35}%
        }
        \put(70,40){%
            \input{tikzfigures/validation_unit_fstudy_1.25}%
        }
    \end{overpic}
    \caption{}
    \label{fig: error_vs_force}
\end{subfigure}
    \begin{subfigure}[t]{0.9\textwidth}
        \centering
        \begin{overpic}[width=\textwidth,height=0.5\textheight,keepaspectratio]{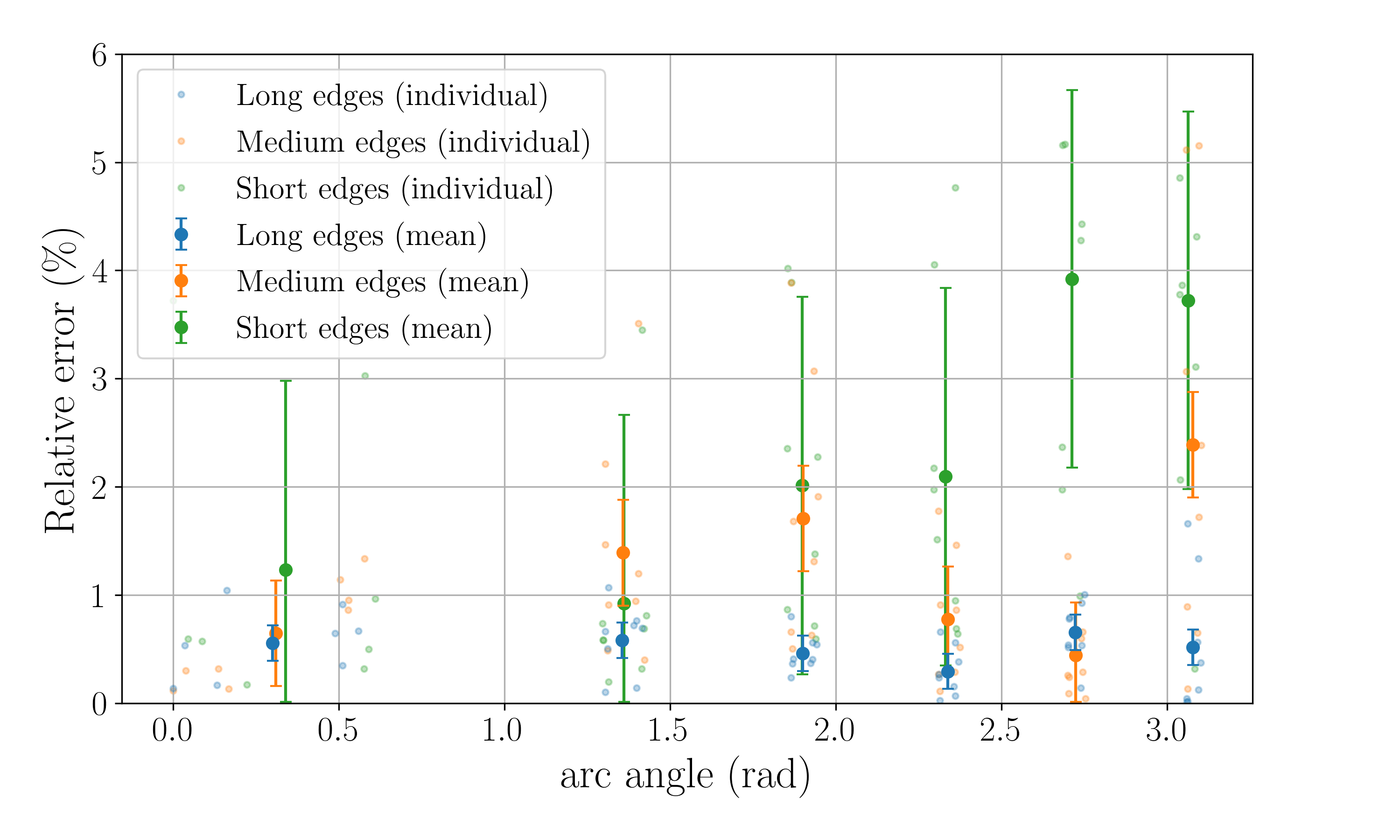}
        \put(20,20.5){%
            \input{tikzfigures/validation_unit_ls1.00}%
        }
        \put(85,15){%
            \input{tikzfigures/validation_unit_ls0.65}%
        }
    \end{overpic}
        \caption{}
        \label{fig: error_vs_theta}
    \end{subfigure}
    \caption{a) Blue axis (left): Relative edge length error versus maximum specimen stress. Orange axis (right): Average measured stress–strain relationship of TPU. High accuracy is observed for stresses up to 7.3 MPa. b) Relative edge length error versus arc angle for Long ($\approx$62mm), Medium ($\approx$21mm), and Short ($\approx$5.8mm) edges.}
    \label{fig:limitations}
\end{figure}

Expected limitations of the proposed methods are discussed in \cref{sec:limitations-methods}. The performance of these methods can be quantified using \cref{eq:score}. In \cref{fig: error_vs_force}, the edge length error is plotted as a function of the edge stress to identify the method’s limits. The results show that the methods remain accurate for stresses up to 7.3 MPa, with errors below 1\%. Reduced accuracy at higher stresses is attributed to the sensitivity of the stress–strain curve in this range, where small strain offsets produce larger errors.

Next, the edge length error is plotted against the edge arc radius for Long ($\approx$62 mm), Medium ($\approx$21 mm), and Short ($\approx$5.8 mm) edges, as shown in \cref{fig: error_vs_theta}. The data indicate that the methods maintain accuracy even at very high arc angles. Still, two sources of increased error are observed: short edges alone cause higher errors, and the combination of short edges with high arc angles leads to even larger errors. Both effects are consistent with the limitations discussed in \cref{sec:limitations-methods}. Notably, accuracy remains within 2.1\%, even for short edges, provided that edges are printed with an arc length below 2.4 rad.
Environmental conditions during printing and testing were consistent (Humidity: $\mu = 25.3\%$, $\sigma^2 = 0.51\%\!^2$; Temperature: $\mu = 23.0\,^\circ\mathrm{C}$, $\sigma^2 = 0.26\,^\circ\mathrm{C}^2$).

\section{Conclusion}\label{sec:conclusions}
The presented work demonstrated a scalable and accessible approach for fabricating network structures with programmable tension gradients using standard FFF techniques. By introducing a design algorithm that transforms tensioned 2D, 2.5D, and 3D cable networks into flat, relaxed layouts, the method enabled the direct 3D printing of entire cable networks as single, continuous pieces. This innovation addressed the longstanding challenge of achieving precise, pre-programmed cable tensions in miniaturized tensegrity and network structures---an essential factor for their mechanical performance and functional adaptability. The approach further streamlined the manufacturing process by minimizing assembly steps and errors, and it leveraged numerical optimization and geometric transformations to ensure that printed networks, upon suspension, realized their intended tension distributions and structural forms.

Experimental validations, including the fabrication of spider web-inspired networks, moment-exerting meshes, and a classic tensegrity structure, highlighted the method’s versatility in reproducing complex tension gradients. Validation steps confirmed that the printed networks closely matched their designed geometries and mechanical properties, with average edge strain errors remaining low: the methods remained accurate within 1\% for tensile stresses up to 7.3 MPa, and within 2.1\% for edges as short as 5.8 mm, provided the arc angle of short edges does not exceed 2.4 rad. While some limitations persisted---such as restrictions on manufacturable tension gradients and geometric configurations due to material and process constraints---the method’s compatibility with widely available 3D printers democratizes access to programmable tensegrity fabrication. This opens new avenues for customizable, lightweight, and adaptive devices in fields ranging from medical orthotics to wearable robotics, paving the way for a broader adoption of tension-programmed structures in both research and practical applications.\\

%
%


\clearpage
\appendix
\section{Constitutive model} \label{AP:material model}
\begin{figure}
    \centering
    \includegraphics[width=0.9\linewidth]{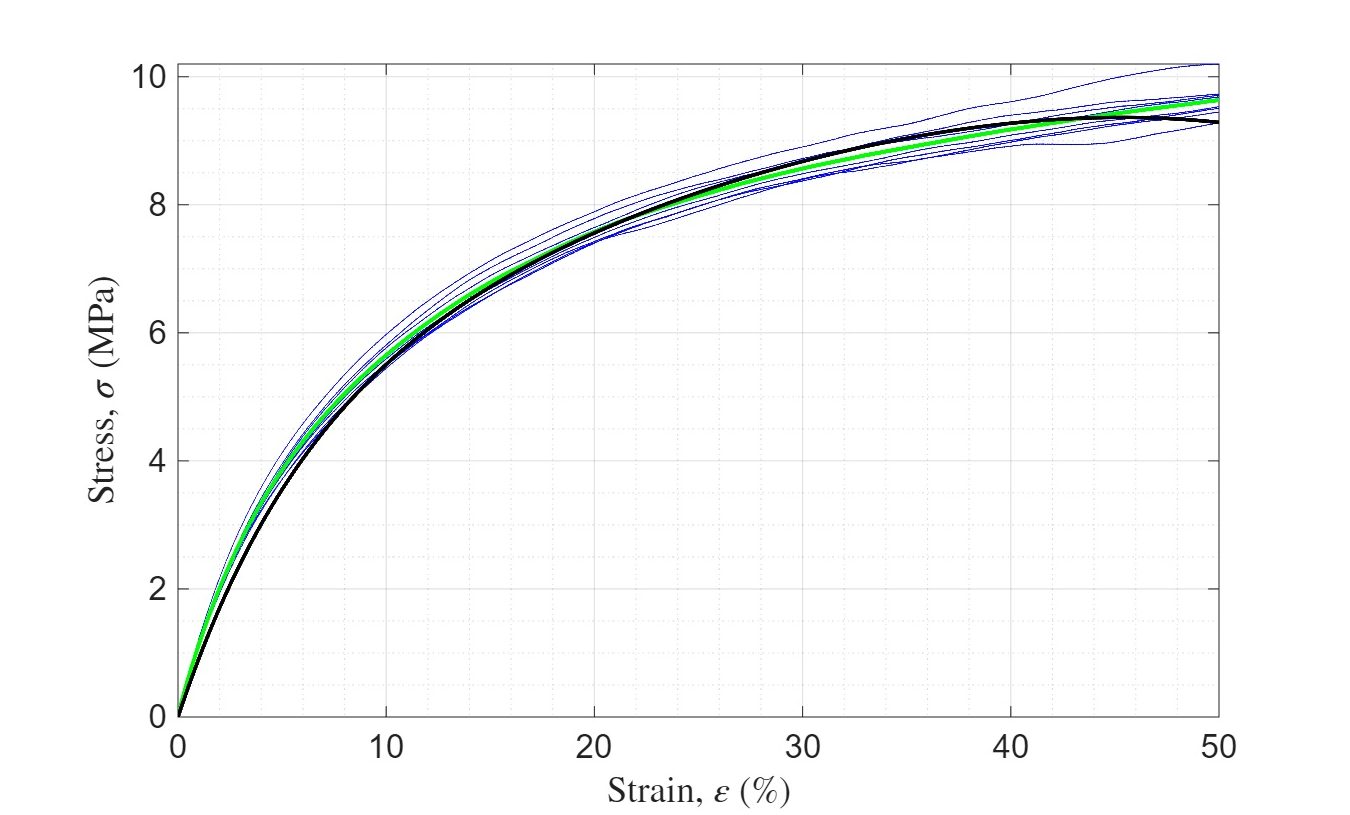}
    \caption{Stress-strain curve of Overture TPU. The blue curves are experimental data for eight specimens. The green curve is the average of the eight specimens. The black curve is the Ogden Material Model.}
    \label{fig:stress-strain}
\end{figure}

To determine the unstretched lengths ($l^0$) of the edges in a structural network, an accurate constitutive model is required. In this study, it was necessary to conduct stress-strain tests to model the nonlinear behavior of the Overture TPU, as shown in \cref{fig:stress-strain}. Following ASTM Standard D882-18, eight specimens with 5$\times$0.2$\times$100 mm$^3$ dimensions were tested at a displacement rate of 10 mm/min. The maximum material strain was 50\%. An Instron 3300 with a 500 N load cell was utilized to conduct uniaxial stress-strain tests.

It is challenging to determine the area of a sheet of TPU only one 3D print layer thick. However, calculating the stress using the theoretical cross-sectional area from the CAD model is insufficient to achieve a reasonably accurate stress-strain curve. Therefore, the cross-sectional area of a single edge $A$ in an FFF network is determined by printing a single edge of length $L_e$ = 36.22 m, with negligible tolerances. After printing, the length of the used filament was $L_f = 1.62 \pm0.02$ m. The cross-sectional area of the filament $A_f$ was specified in the technical datasheet $A_f = 1.75 $mm$^2\pm$ 0.02. Using conservation of volume, the cross-sectional area of a single edge was determined according to:

\begin{equation}
    A = \left(A_f \cdot L_f \right)/ L_e = 0.0783 \pm0.002 \text{ mm}^2
\end{equation}

Typically, a constitutive model for a nonlinear elastic material is based on the strain energy density function $S$. Thus, the stress can be obtained simply by taking the derivative of $S$ with respect to strain. Characterizing the properties of rubbery materials is often based on the stretch ratio rather than strain; i.e. $\lambda=l/l^o=1+\varepsilon$. Because the deformation in TPU is three-dimensional, the strain can be related to the principal stretch ratios \cite{dal2020extended}: 

\[
I_1=\lambda_1^2+\lambda_2^2+\lambda_3^2, \hspace{10mm} 
I_2=\lambda_1^2 \lambda_2^2+\lambda_2^2 \lambda_3^2 +\lambda_3^2 \lambda_1^2, \hspace{3 mm} \text{and} \hspace{5mm} 
I_1=\lambda_1^2 \lambda_2^2 \lambda_3^2
\]

Where $\lambda_i$ denote the principal stretches in the 1-2-3 (or xyz) directions. The Ogden model was utilized in this study, as it provided a good correlation with the experimental data in \cref{fig:stress-strain}. The strain energy function for this model is \cite{durna2024hyper}: 

\begin{equation}
    S=\sum_{i} \frac{\mu_i}{\alpha_i}\Big(\lambda_{1}^{\alpha i}+\lambda_2^{\alpha i}+\lambda_3^{\alpha i}-3\Big) 
\end{equation}

The $\alpha_i$ and $\mu_i$ are material constants estimated numerically using Hyper-Data, a Matlab-based optimization \cite{durna2024hyper} based on the uniaxial test data. These constants were $\alpha_i = (0.0024, 7.04, -13.6)$ and $\mu_i = (81634, -5.64, -6.26)$. The material response was assumed to be incompressible and isothermal, hence $\lambda_{1} \lambda_{2} \lambda_{3}=1$. This assumption was used to obtain an expression for $ \lambda_{2}$ and $\lambda_{3}$ using only uniaxial stress-strain tests, as follows \cite{wiersinga2022hybrid}: 

\begin{equation}
    \lambda=\lambda_{1}=\frac{l}{l_{o}}, \lambda_{2}=\lambda_{3}=\sqrt{\frac{l_{o}}{l}}
\end{equation}


Finally, the uniaxial nominal stress was obtained by differentiating $S$ with respect to $\lambda$ instead of $\varepsilon$. 

\begin{equation}
    \sigma=\sum_{i}\mu_{i} \left(\lambda_{1}^{\alpha_{i}-1}-\lambda_{1}^{\frac{-\alpha_{1}}{2}-1}\right)
\end{equation}

\section{Dynamic Loading Behavior}\label{AP:hysteresis}

\begin{figure}
    \centering
    \includegraphics[width=0.9\linewidth]{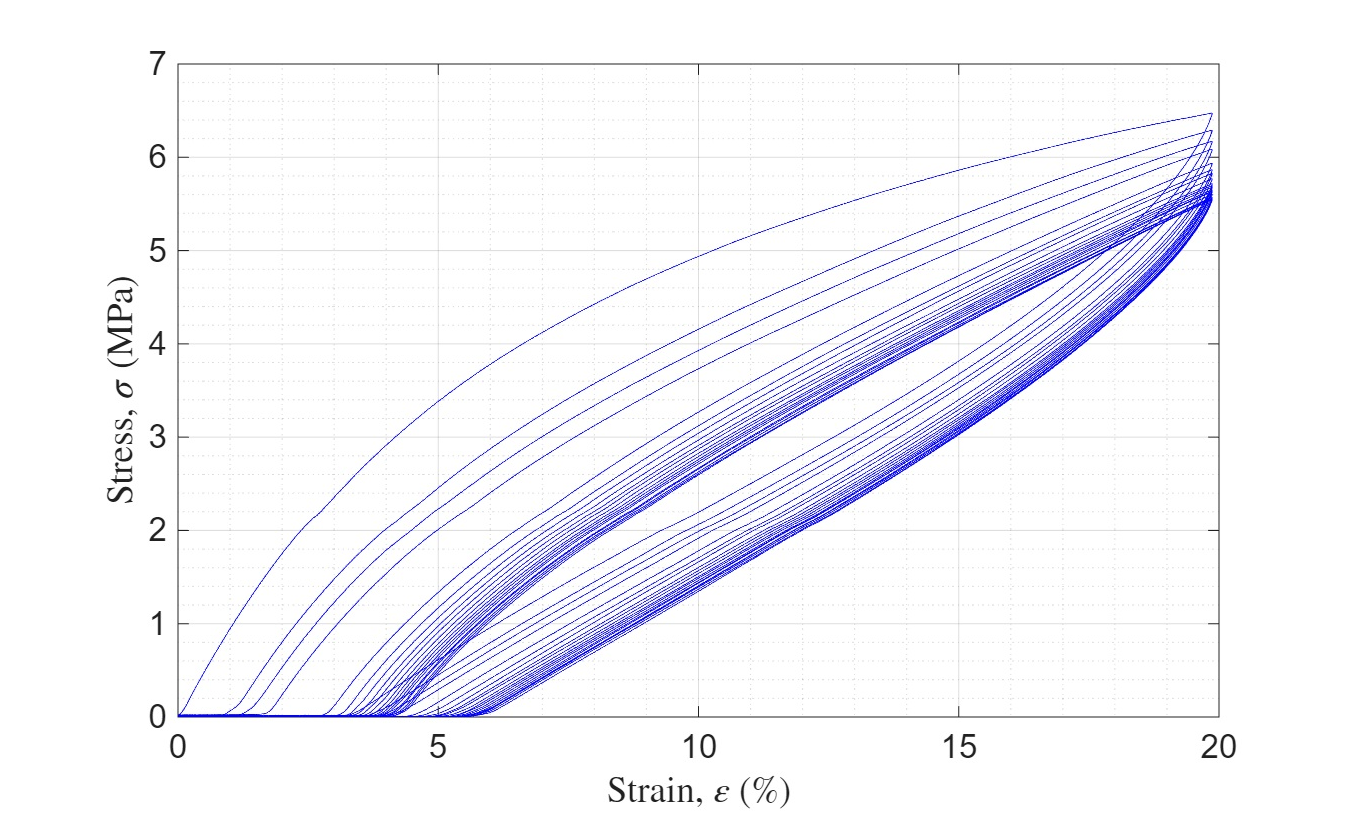}
    \caption{Average stress-strain curve for 3 specimens of Overture TPU.}
    \label{fig:hystersis}
\end{figure}

\begin{figure}
    \centering
    \includegraphics[width=0.8\linewidth]{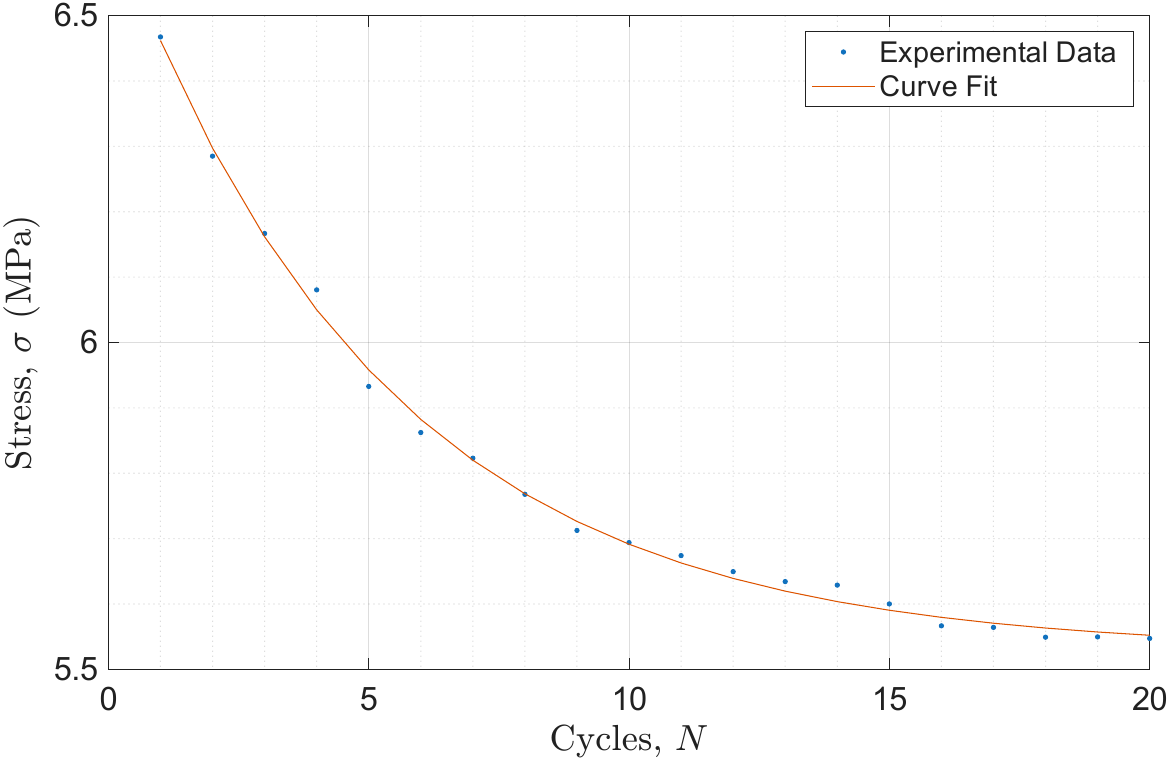}
    \caption{Exponential Decay of the stress at 20\% strain as a function of the number of cycles.}
    \label{fig:ExpDecay}
\end{figure}

For control scenarios in future applications, it is desirable to predict how TPU cables will behave under dynamic loading conditions. Hysteresis testing was performed to measure the differences in the stress-strain curves for loading vs. unloading. Cyclic testing was performed to determine how the stress-strain curve would change from cycle to cycle. For these tests, each specimen was loaded to 20\% strain and, without pausing, returned to zero strain, at which point it was allowed to recover until stress stabilized. This process was performed 3 times. After the 3rd cycle recovery, the process was repeated 17 times for a total of 20 cycles, except that the specimen was not allowed to recover between cycles. Loading and unloading were performed at a rate of 10 mm/min. An exponential decay rate was observed (see \cref{fig:hystersis}). A curve fit of the exponential decay of the stress at 20\% strain yields $\sim 26$ cycles to settle (see \cref{fig:ExpDecay}).

\begin{figure}
    \centering
    \includegraphics[width=0.8\linewidth]{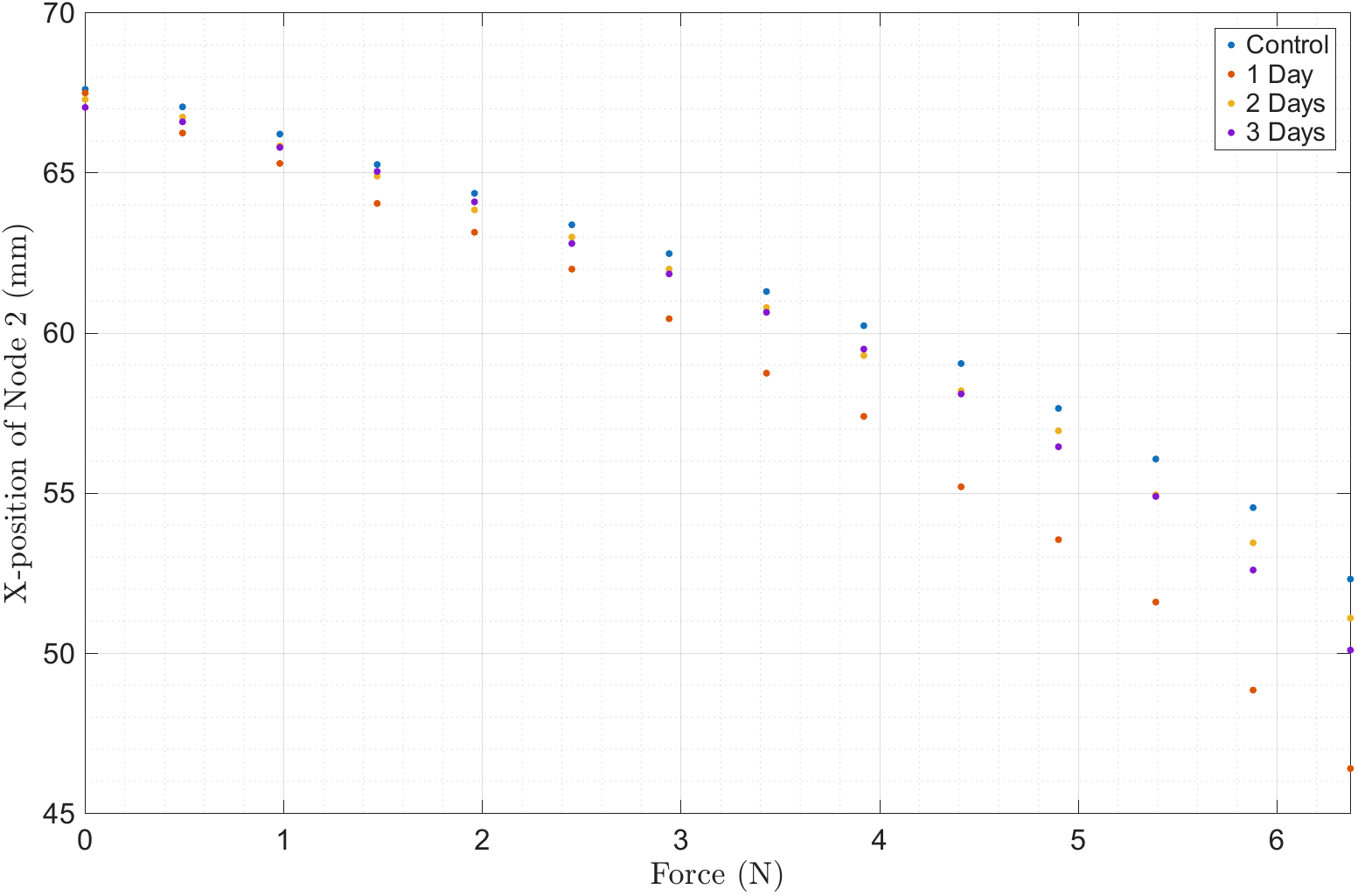}
    \caption{Average Position vs Mass plot of 2 arrays when allowed to rest for 1 day, 2 days, and 3 days.}
    \label{fig:Avg_Daily}
\end{figure}

It was also desirable to ascertain the time constant for the recovery of TPU. Two 1 by 1 tensegrity arrays \cite{brown_development_2024} were tested by applying loads from 0 to 650 grams (0 to 6.37 Newtons) in increments of 50 grams to the control string. The position of Node 2 was then measured. Each array was allowed to rest for 24 hours before being tested again. The process was then repeated with a two- and three-day rest. The results are shown in \cref{fig:Avg_Daily}. It was determined that TPU needs $\sim 2$ days to recover fully.

Strain-rate dependency is another important characteristic of polymers to consider. In this work, we did not consider the loading rate. Instead, we assumed that the test frames were loaded sufficiently slowly to be considered quasi-static.

\section{Additional Methods}
Additional methods are detailed here.
\subsection{Validation methods}\label{Ap:validation_methods}
\begin{figure}[h!]
    \centering
    \includegraphics[width=\linewidth]{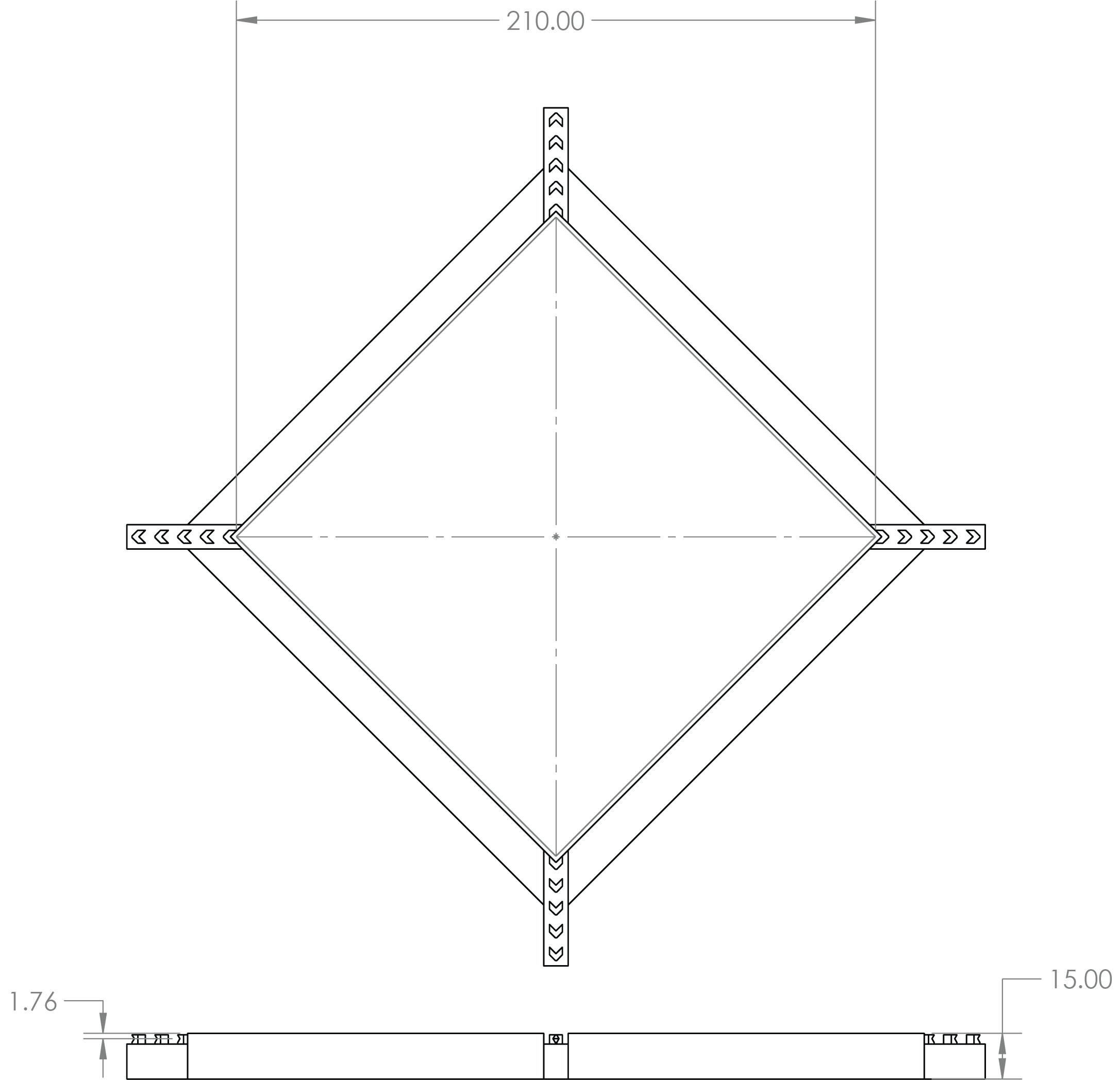}
    \caption{Technical drawing of the Bottom and Side view of the frame used for validating the long-edged unit cells. A similar frame was printed for unit cells with Short and Medium edges, but instead of 210 mm, the diagonal length between fixed points was 70 mm.}
    \label{fig:frame}
\end{figure}

Results of validation were set out in \cref{sec:validation}, where unit cells with long ($\approx$62mm), medium ($\approx$21mm), and short ($\approx$5.8mm) edges were suspended into a frame. The frame was 3D printed from PETG. The unit cells are printed with a 9 mm long loop, allowing them to be suspended in the frames. The hooks on the frame were designed such that the fixed points $\textbf{x}_f$ are located 210 mm apart for the unit cells with long edges, or 70 mm apart for the unit cells with medium and short edges. The average edge length error is calculated using equation \cref{eq:score}. Notably, the edges used for suspending the unit cells are not included in the analysis. The motivation for this is that any offset between the grid line paper and the PETG frame will cause additional error in the connecting edges, but not in the other edges. In \cref{fig:frame} a technical drawing of the frame is depicted.


\subsection{Flattening with cylindrical or spherical coordinates}\label{Ap:flattening_methods}
In order to print curved 2D or 3D networks, the network needs to be flattened (see \cref{sec:flattening}). The detailed approach to flattening using a coordinate transformation is:
\begin{enumerate}
    \item \textbf{Translation:} Shift \(\mathbf{x}\) such that the vertices have the center point \(\mathbf{c}\) as the origin: 
    \[
    \mathbf{p}_j = \mathbf{x}_j - \mathbf{c}
    \]    
    \item \textbf{Rotation(s):} Rotate the points such that the user-defined unit vector \(\mathbf{t}\) aligns with the z-axis.
    \item \textbf{Coordinate transformation:} Apply a coordinate transformation:
    \begin{center}
    \renewcommand{\arraystretch}{1.4}
    \begin{tabular}{>{\raggedright\arraybackslash}m{0.45\textwidth} >{\raggedright\arraybackslash}m{0.45\textwidth}}
    \textbf{Cylindrical coordinates} & \textbf{Spherical coordinates} \\
    \(
    \theta_j = \mathrm{atan2}(p_{j,y}, p_{j,x})
    \) &
    \(
    \theta_j = \mathrm{atan2}(p_{j,y}, p_{j,x})
    \) \\
    \(
    r_j = \sqrt{p_{j,x}^2 + p_{j,y}^2}
    \) &
    \(
    \phi_j = \mathrm{atan2}(p_{j,z}, \sqrt{p_{j,x}^2 + p_{j,y}^2})
    \) \\
    \(
    z_j = p_{j,z}
    \) &
    \(
    r_j = \sqrt{p_{j,x}^2 + p_{j,y}^2 + p_{j,z}^2}
    \)
    \end{tabular}
    \end{center}
    \item \textbf{Dimensional reduction:} The radial coordinates \(\mathbf{r}\) are disregarded, and the angular coordinates \ are scaled with the mean radius \(\bar{r}\), such that for cylindrical coordinates
    \[
    x_j = \theta_j \cdot \bar{r}, \quad y_j = z_j,
    \]
    and for spherical coordinates:
    \[
    x_j = \theta_j \cdot \bar{r}, \quad y_j = \phi_j \cdot \bar{r}.
    \]
\end{enumerate}
\subsection{Determining arc parameters} \label{Ap:arc_params}
The interpolation function is only set up for a feasible space. Lengths are positive and an arc length is always longer than its chord length, i.e.  $0\leq l^1/l^0\leq1$. The limits for $\alpha$ become $0\leq \alpha < \pi$ when only allowing minor arcs and positive angles. The arc radius is found fast by evaluating $R = l^0/\alpha$.
\begin{figure}[h]
    \centering
    \includegraphics[width=.7\linewidth]{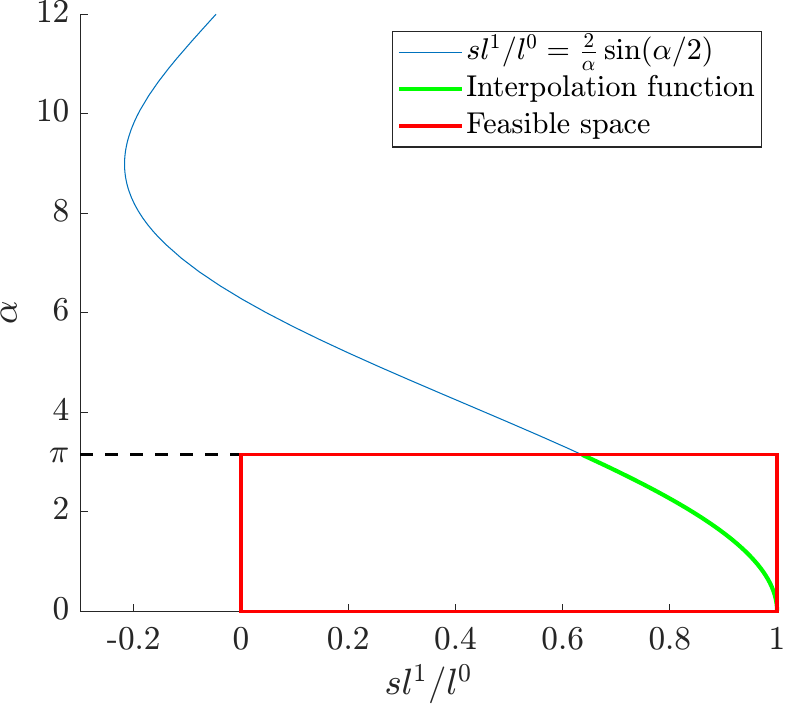} %
    \begin{tikzpicture}[overlay, remember picture]
        \begin{scope}[shift={(-2.5cm, 4cm)}] 
            \draw[thick] (0,0) -- (60:3);
            \draw[thick] (0,0) -- (120:3) node[midway, below left] {$R$};
            \draw[thick] (60:3) arc[start angle=60,end angle=120,radius=3] node[midway, above] {$l^0$};
            \draw[dashed] (60:3) -- (120:3) node[midway, below] {$sl^1$};
            \draw[thick] (120:0.5) arc[start angle=120, end angle=60, radius=0.5] node[midway, above] {$\theta$};
        \end{scope}
    \end{tikzpicture}
    \caption{Determining arc parameters. $\alpha$ can be determined fast by evaluating a cubic interpolation function in the feasible space. }
    \label{fig:arc-parameters}
\end{figure}

\section*{Acknowledgment}
The authors are grateful for support provided by the National Science Foundation via CAREER Award No. 2440838: Harnessing the Dynamic Functions of Irregular Structural Patterns. The authors will also  acknowledge Victoria Leigh's and Keira Morrissey's extensive efforts to help validate the proposed methods. The authors acknowledge the author of the GitHub repository GcodeGenerator, Tibor Bar\v{s}i,  which was used to generate the 3D printer machine code of all networks in this paper. The authors thank Reagan Rogers and Myah Derderian for their help in tensile testing the TPU. The authors thank Christy Graves for designing the graphical abstract.

\section*{Conflict of interest}
The authors declare they have no conflict of interest.

\printbibliography

\end{document}